\documentclass{statsoc}

\pdfoutput=1 

\usepackage{amsmath,amsfonts,amssymb}
\usepackage{latexsym,textcomp,cancel}
\usepackage{graphicx}
\usepackage{natbib}
\usepackage{appendix}
\usepackage{url,color}
\usepackage{multirow, rotating}
\usepackage{cancel}
\usepackage{subfigure}

\usepackage{fullpage}

\newcommand{\bfxi} {\boldsymbol{\xi}}
\newcommand{\bfmu} {\boldsymbol{\mu}}

\newcommand{\bfzeta} {\boldsymbol{\zeta}}

\newcommand{\bfpi} {\boldsymbol{\pi}}

\newcommand{\bfTheta} {\boldsymbol{\Theta}}
\newcommand{\bfSigma} {\boldsymbol{\Sigma}}
\newcommand{\bfOmega} {\boldsymbol{\Omega}}

\newcommand{\bfPi} {\boldsymbol{\Pi}}

\newcommand{\bfw} {\mathbf{w}}

\newcommand{\bfx} {\mathbf{x}}

\newcommand{\bfY} {\mathbf{Y}}

\renewcommand{\Pr}{\mathsf{Pr}}
\newcommand{\E}{\mathsf{E}}

\newcommand{\normal}{\mathsf{Normal}}

\newcommand{\Dir}{\mathsf{Dir}}

\newcommand{\IWis}{\mathsf{IW}}

\newcommand{\bet}{\mathsf{Beta}}

\newcommand{\Ber}{\mathsf{Ber}}

\title[Modeling and prediction of financial trading networks] {Modelling and prediction of financial trading networks:  An application to the NYMEX natural gas futures market}

\author[Betancourt {\it et al.}]{Brenda Betancourt} \email{bbetanc1@soe.ucsc.edu}
\author{Abel Rodr\'{\i}guez}
\address{University of California, Santa Cruz, U.S.A.}
\author{Naomi Boyd}
\address{West Virginia University}

\begin{document}

\begin{abstract}
Over the last few years there has been a growing interest in using financial trading networks to understand the microstructure of financial markets.  Most of the methodologies developed so far for this purpose have been based on the study of descriptive summaries of the networks such as the average node degree and the clustering coefficient.  In contrast, this paper develops novel statistical methods for modeling sequences of financial trading networks.  Our approach uses a stochastic blockmodel to describe the structure of the network during each period, and then links multiple time periods using a hidden Markov model.  This structure allows us to identify events that affect the structure of the market and make accurate short-term prediction of future transactions.  The methodology is illustrated using data from the NYMEX natural gas futures market from January 2005 to December 2008.
\end{abstract}

\section{Introduction}\label{sec:intro}

Financial trading networks are directed graphs in which nodes correspond to traders participating in a financial market, and edges represent pairwise buy-sell transactions among them that occurr within a period of time.  Financial trading networks contain important information about patterns of order execution in order-driven markets; hence, they provide insights into aspects of market microstructure such as market frictions, trading strategies, and systemic risks.

For example, consider the role of financial trading networks in understanding the effect of market frictions on market microstructure.  In the absence of market frictions, we could expect orders from different traders to be matched randomly.  However, real trading networks often exhibit features such as elevated transitivity or preferential attachment among certain groups of actors \citep{AdBrHaKi10}, which are inconsistent with random matching.  In the case of open-outcry markets, these features can be partially explained by sociological factors (for example, see \citealp{Za04}). Alternative explanations include the effect of different market roles (e.g., liquidity providers/takers) or trading strategies (e.g., long vs.\ short strategies), see \citealp{OzWaYaBi10} or \citealp{HaKoNiOsWe12}.

Financial trading networks also provide information that is key in the assessment of systemic risks.  Analysis of the evolution of financial trading networks can aid in tests of financial market stability (or fragility as it may be) by financial regulators to ensure that events such as a large trader failures do not serve to destabilize financial markets.  For example, in the event of a large trader failure, an understanding of their network will help guide regulators through the process of unwinding their positions and may dictate whether those positions are unwound in the open market or through a transfer to a suitable counterparty \citep{BoHaNo11}.  Financial trading networks can also be used to identify important traders that play a critical role in the market (for example, by acting as de facto market makers or liquidity providers).  In addition, they can also help us identify frequent counterparties of specific traders which may aid in regulatory oversight by federal agencies and market exchanges alike.  Indeed, there is evidence that price distortion and manipulation may be more likely between frequent counterparties than by one agent acting in isolation \citep{HaChSc94}.


The literature on the mathematical modeling of financial trading networks is limited.  Theoretical approaches that explain the structure of a financial network as the outcome of a game have recently been developed (e.g., see \citealp{OzWaYaBi10} and \citealp{HaKoNiOsWe12}), but they are of limited practical applicability.  Most of the empirical work on trading networks has focused on the use of summary statistics such as degree distributions, average betweenness and clustering coefficients \citep{Newman,AdBrHaKi10}. These type of approaches provide some interesting insights into market microstructure, but suffer from two main drawbacks.  First, the summary statistics to be monitored need to be carefully chosen to ensure that relevant features of the market are captured (for an example of this, see Section \ref{se:data}).  Although some of the game-theoretic work mentioned before might provide insights into which network summaries should be monitored, the choice is typically ad-hoc and the selection is often incomplete.  Second, and more importantly, approaches of this type are not helpful in predicting future interactions among traders.

In this paper we move beyond descriptive network summaries to focus on stochastic models for array-valued data that place a probability distribution on the full network.  The simplest such model is the class Erd\"os-R\'enyi model \citep{ErRe59}, which assumes that interactions among any two traders occur independently and with constant probability that is independent of the identity of the traders.  This class of models, although well studied from a theoretical perspective, is too simplistic to accommodate most realistic networks.  As an alternative, \citet{Frank} proposed the class of Exponential Random Graph models (ERGMs), also called $p^*$ models.  These models formalize the use of summary statistics by including them as sufficient statistics in exponential-family models. A temporal version of the ERGM was introduced in \citet{HanFuXi10} and further developed in \citet{CranDes11} and \citet{SnijSteBunt10}.  The class of $p_1$ models, which extends generalized linear models to array-valued data, was originally proposed by \citet{Holland} and extended to dynamic settings in \cite{BaCa96}, \cite{GoldZhAi09} and \cite{Kolac09}.  Another related approach was introduced in \citet{Hoff2} using the concept of latent social space models.  In this class of models the probability of a link between nodes increases as they occupy closer positions in latent social space.  Models based on latent social spaces have again been extend to dynamic settings by \citet{SarkarMoore05} and \citet{SewChen15}, among others.

The model discuss in this paper extends the class of stochastic blockmodels first introduced in \citet{Wang} to account for time dependence.  Stochastic blockmodels rely on the concept of structural equivalence to identify groups of traders (which we shall refer to as \textit{trading communities} in the context of this application) with similar interaction patterns.  Model-based stochastic blockmodels have been developed as array-valued extensions of traditional mixture models, and dynamic versions of these models have been recently proposed. For example, \citet{Nowicki} presented a simple Bayesian model that uses a finite mixture model and a Dirichlet prior for the probabilities of the latent classes. A dynamic variant with a first order Markov model is presented in \citet {YangChiZhu11}. An extension of this model that relies on infinite mixture models based on the Dirichlet process have been proposed by \citet{Kemp} and \citet{Xu}. More recently, \citet{Airoldi} introduced the idea of mixed membership stochastic blockmodels for binary networks wherein the actors can belong to more than one latent class to explore subjects with multiple roles in the network. The work of \citet{XiFuSo10} develops its temporal extension.

Other approaches to dynamic stochastic blockmodels include the work of \citet{WangTang14} who proposed a method for change-point detection using hypothesis testing and locality statistics to identify anomalies over time, and the state-space model of \citet{SuHero14} which introduces the extended Kalman algorithm as an alternative to MCMC. An extensive review of methods for anomaly detection in dynamic etworks is presented in \citet{RanShiKou15} including some relevant probabilistic models. For example, \citet{HeaWePlat10} utilizes Bayesian discrete time counting processes with conditionally independent increments in order to identify nodes whose relationships have changed over time (see also \citealp{RobPriebe13}), and \citet{PerryWolfe13} consider a multivariate point process to model directed interactions between actors in continuous time, and explore the impact of homophily and network effects in the prediction of future interactions.

In this paper we propose modeling the dynamics of financial trading networks using an extension of the Bayesian infinite-dimensional model of \citet{Kemp}. The model we propose accounts for dependence of the network structure over time and incorporates more general hierarchical priors on the interaction probabilities as well as the partition structure.  To account for changes in market microstructure over time, the blockmodels associated with different time periods are linked through a hidden Markov model. In finance, regime switching models have been used in many contexts such as applications to model stock returns \citep{GuiTim05,KiMoNe01,PeTim00}, in asset allocation \citep{AnBe02a}, business cycles \citep{Fil94}, and interest rates \citep{AnBe02b}.  As we show in our illustration, by developing a dynamic, fully probabilistic model for array-valued data we are able to monitor structural changes in market microstructure while at the same time making more accurate short-term predictions of future trading patterns.  

\section{Data }\label{se:data}

\begin{figure}
\begin{center}
\includegraphics[scale=0.65,keepaspectratio]{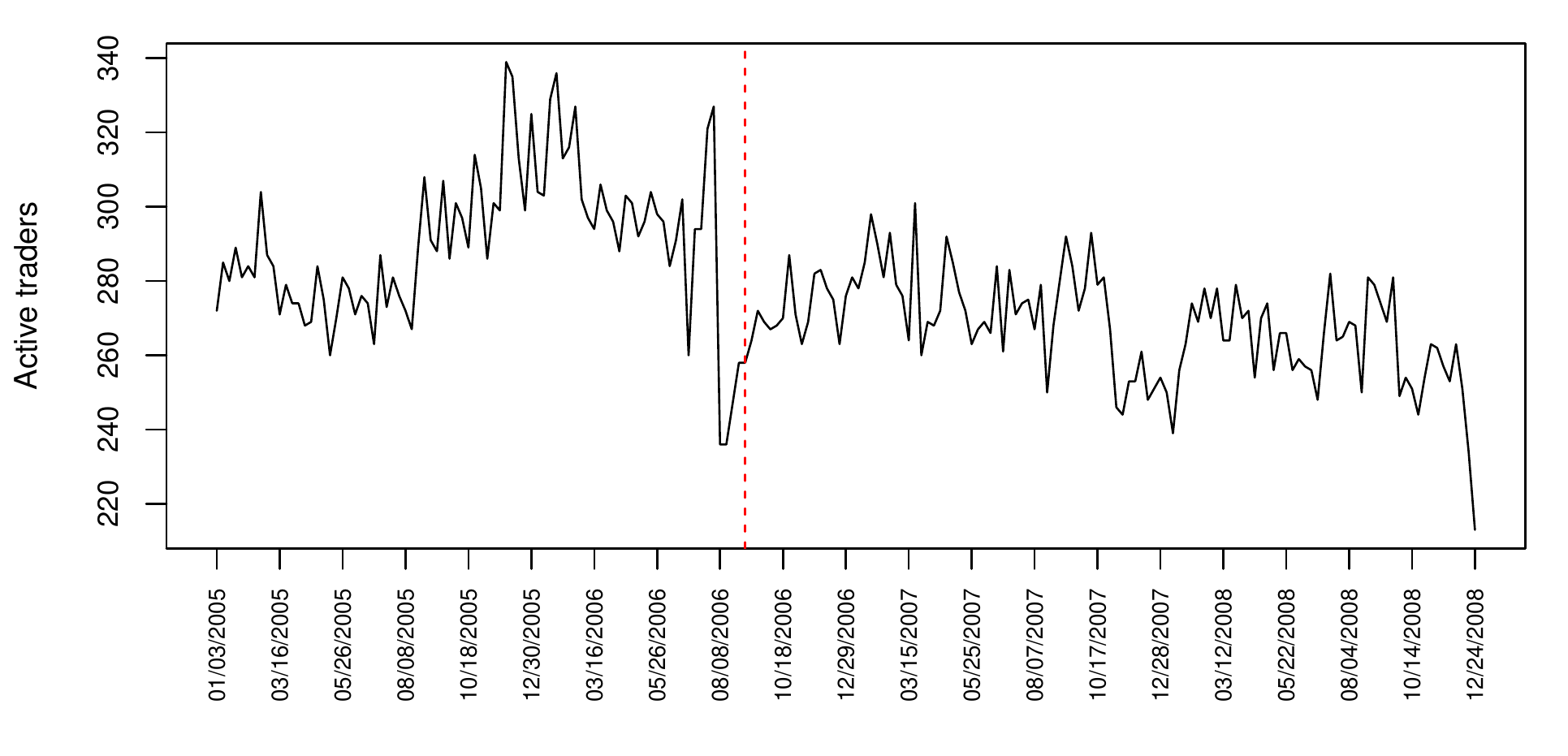}
\end{center}
\vspace{-0.5cm}
\caption{Number of active traders each week in the NYMEX natural gas future market between January 2005 and December 2008.  The dashed vertical line marks September 5, 2006, when electronic trading was introduced in this market.}\label{fi:numtraders}
\end{figure}

The data we analyze in this paper consists of proprietary transactions made by traders in the New York Mercantile Exchange (NYMEX) natural gas futures market between January 2005 and December 2008.  A total of 970 unique traders participate in proprietary transactions at least once over the four years to December 2008.  However, this list includes traders that either abandoned proprietary trading or went bankrupt during the period under study, as well as traders that entered the market after January 2005.  Indeed, only between 240 and 340 traders participated in trades each week (see Figure \ref{fi:numtraders}).  Since we have no detailed information about the times at which different traders entered or left the market, our analysis focuses on 71 traders we identified as being present in the market (although not necessarily active) during the whole period. Note that traders were anonymized and are identified in the paper using numbers.

We used the transaction data to construct weekly trading networks where a link from trader A to trader B was established if there was at least one transaction during that week in which A was the seller and B was the buyer.  Data was grouped weekly because this is a low liquidity market in which daily transactions do not provide any strong signal of community behavior and the number of daily participants is too low compared to the total number of traders involved in the market over the four-year period. Monthly transactions are coarser than weekly transactions but show similar patterns and weekly observations allow us to have a more refined exploration of the network data.  On the other hand, we considered binary networks instead weighted networks (e.g number or volume of transactions) because the presence/absence of links provides enough information to understand the dynamic of the network in terms of traders partnerships and community patterns.  

\begin{figure}
\begin{center}
\subfigure[Mean total degree]{\includegraphics[scale=0.50,keepaspectratio]{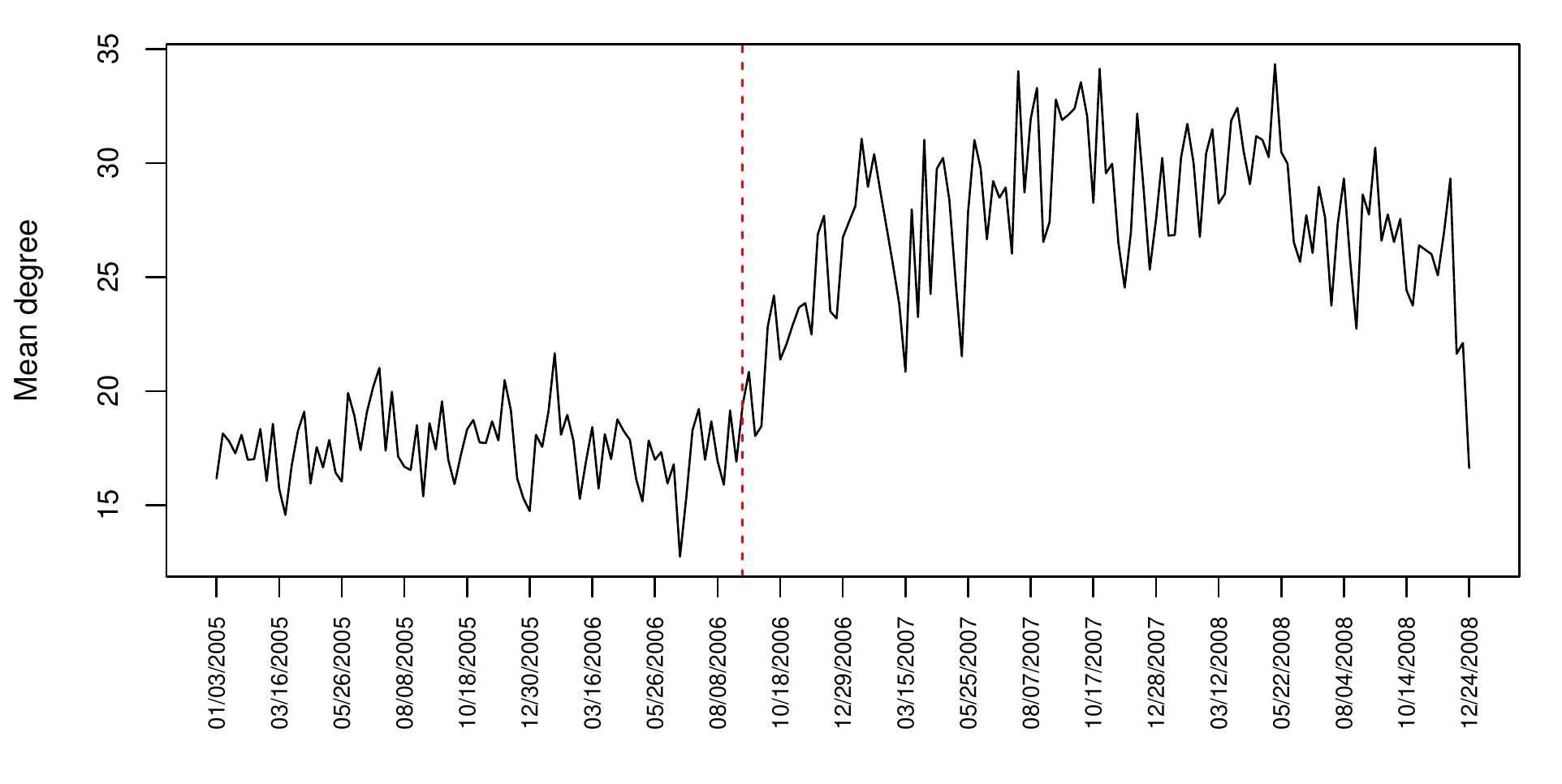}}
\subfigure[Clustering coefficient]{\includegraphics[scale=0.50,keepaspectratio]{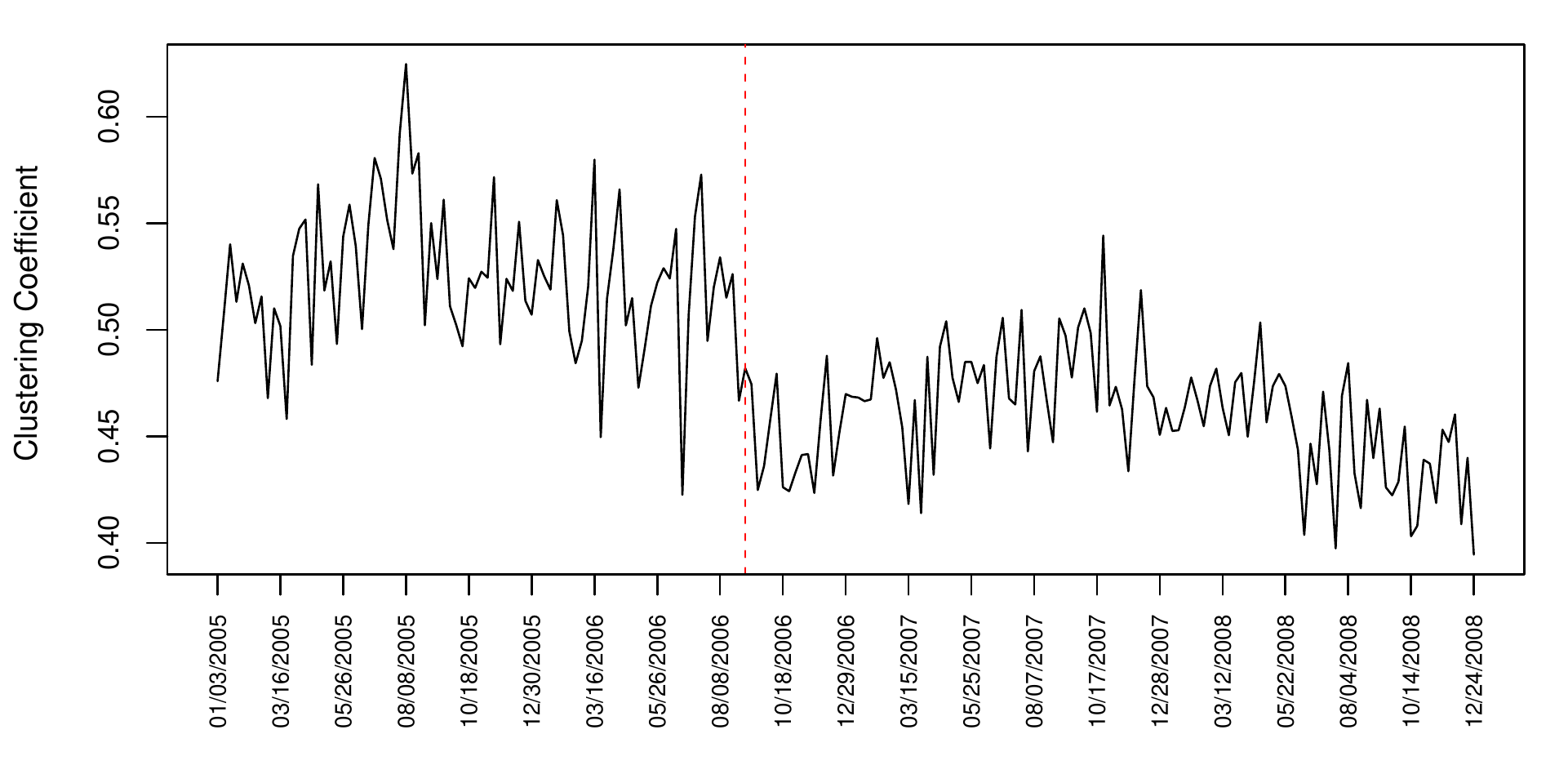}}
\subfigure[Assortativity by degree]{\includegraphics[scale=0.50,keepaspectratio]{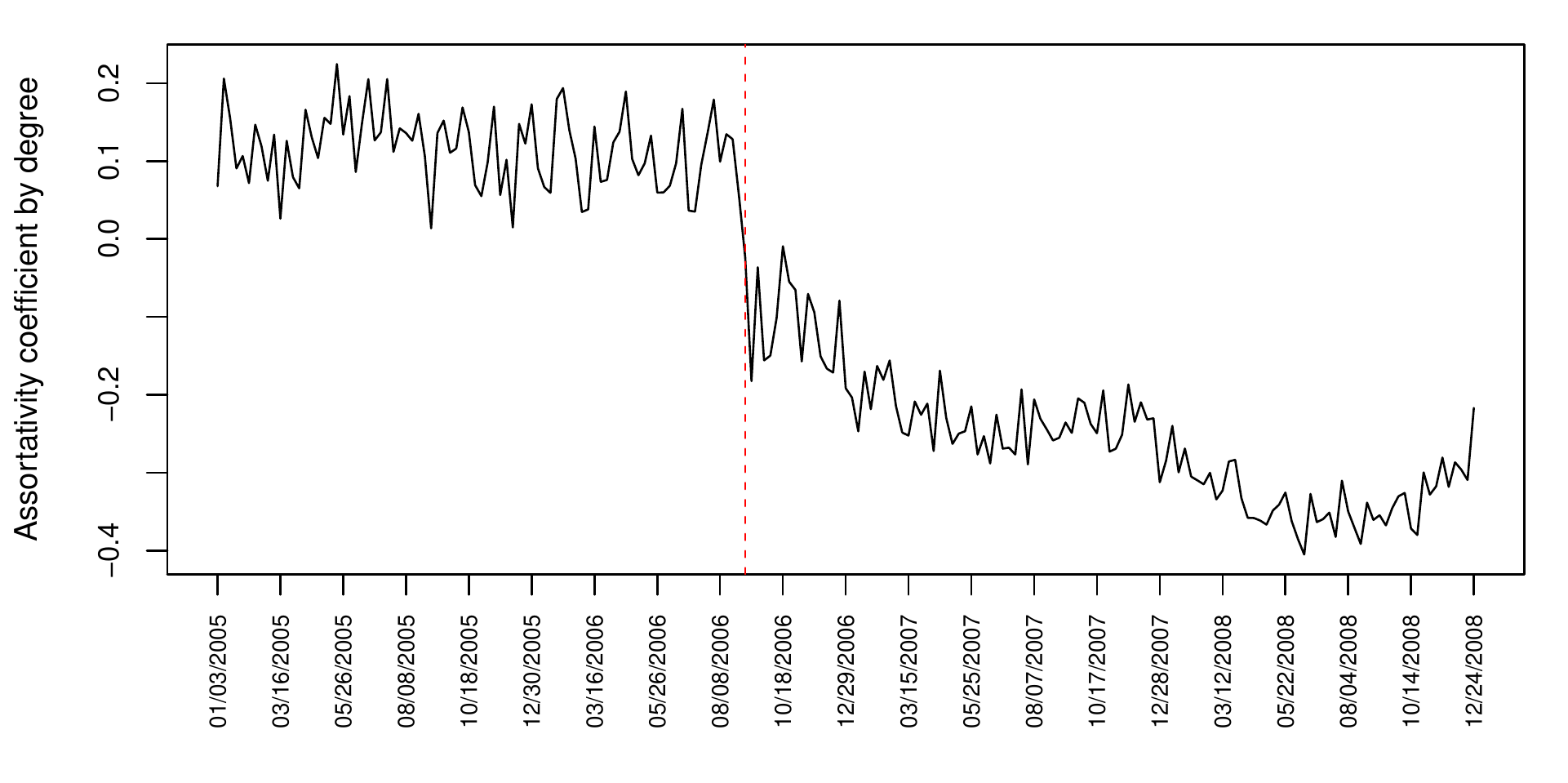}}
\end{center}
\caption{Mean total degree, clustering and assortativity coefficients for weekly trading networks in the NYMEX natural gas futures market between January 2005 and December 2008.  The dashed vertical line marks September 5, 2006, when electronic trading was introduced in this market.}\label{fi:clustcoeff}
\end{figure}

\begin{figure}
\begin{center}
{\includegraphics[scale=0.34,keepaspectratio]{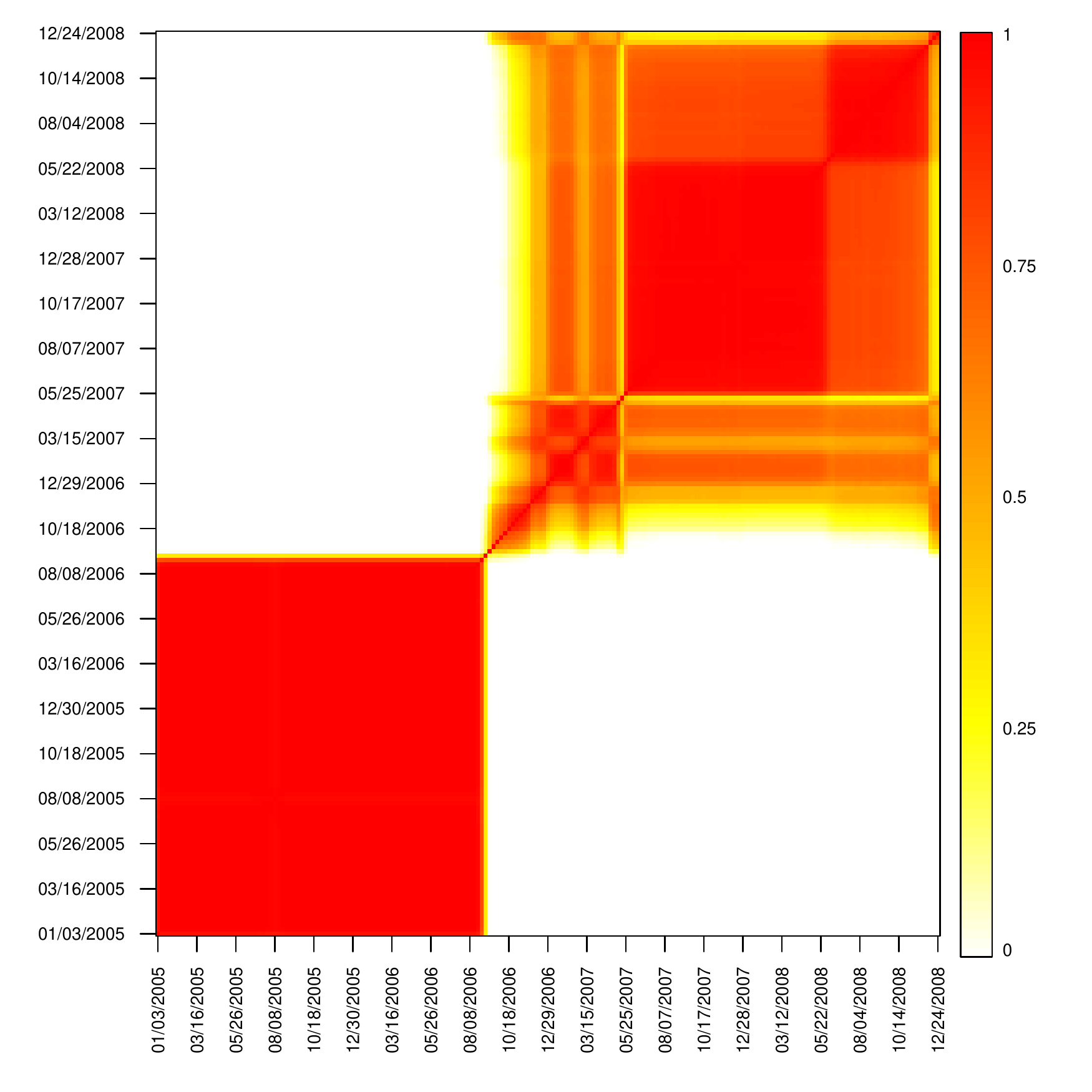}}
{\includegraphics[scale=0.34,keepaspectratio]{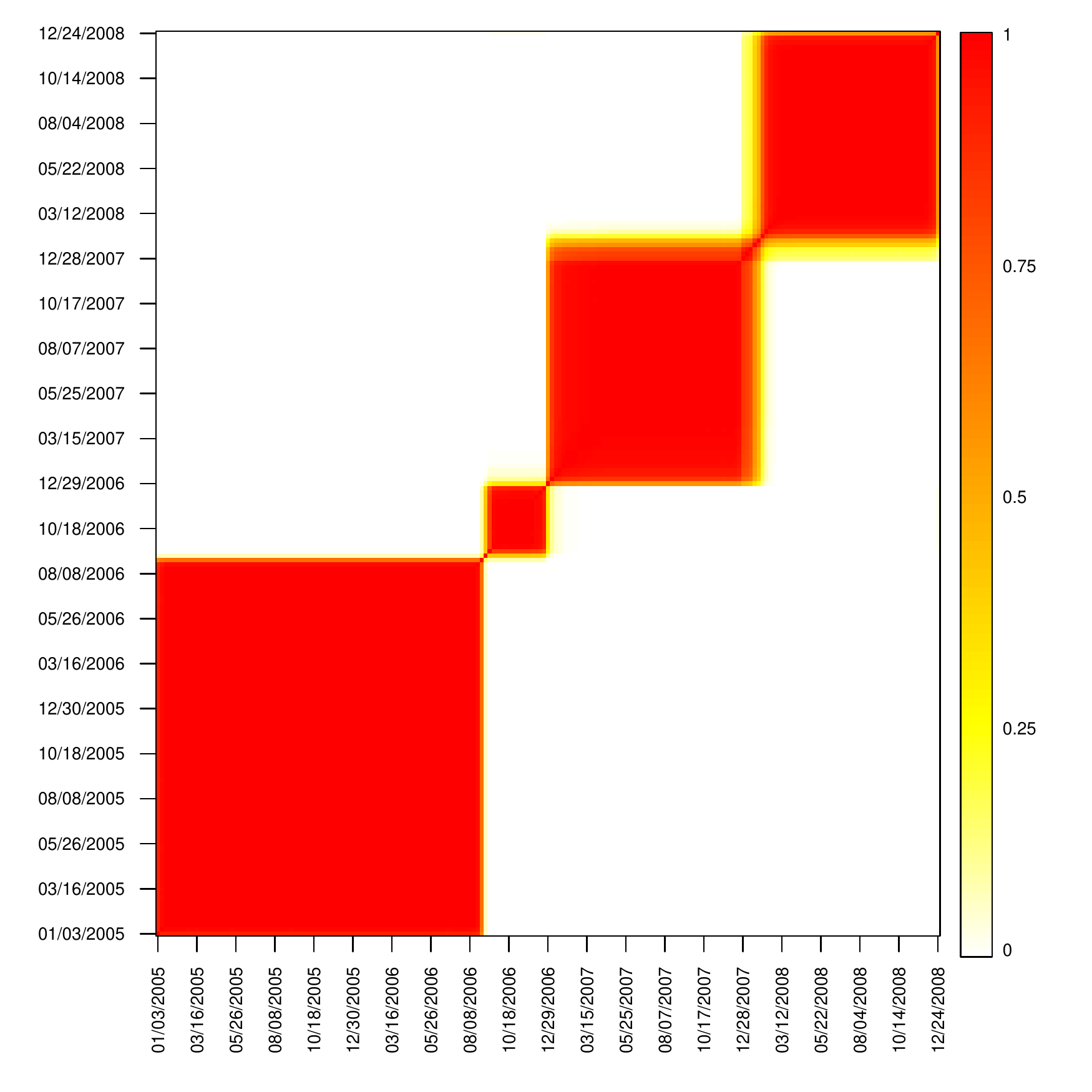}}
\end{center}
\caption{Mean posterior pairwise incidence matrix for the NYMEX networks under a simple hidden Markov model with Gaussian emissions.  The left panel shows the results based on the mean total degree and the clustering coefficient, while the right panel shows the results based on the clustering and assortativity coefficients.}\label{se:incidence_simpleHMM}
\end{figure}

Table \ref{ta:Summary} presents summary statistics for the number of active nodes, mean and maximum in and out degree, the degree correlation, the clustering coefficient and the probability of a link computed over the 201 observed networks for the 71 selected traders. The mean and maximum values of in and out degree are relatively low showing sparsity in the network. The high positive values of the degree correlation suggest that traders tend to make buy and sell transactions with the same partners reflecting high reciprocity in the network. In addition, the clustering coefficient is high compared to the probability of a link suggesting high transitivity in the network and confirming the presence of social network patterns in this data.
\begin{table}
\caption{\label{ta:Summary} Summary of descriptive statistics over 201 weekly trading networks.}
\begin{tabular}{lccccc}
   &{\bf Min}  &{\bf 25\%} &{\bf 50\%}  &{\bf 75\%}  &{\bf Max} \\
\hline \hline
{\bf Active Traders}   &  56 &  66 &  69 & 70 & 71\\
{\bf Mean degree}    & 13   & 18 &  23 & 28 & 34 \\  
{\bf Max. in-degree}    & 19   & 30 &  44 & 54 & 64 \\     
{\bf Max. out-degree}    & 21   & 29 &  48 & 56 & 66 \\     
{\bf Degree correlation}    & 0.838   & 0.903 &  0.923 & 0.939 & 0.966 \\
{\bf Clustering coefficient} & 0.395 &  0.458 & 0.482 & 0.518 & 0.625\\
{\bf Link probability}    & 0.091    & 0.128 &  0.166 & 0.202 & 0.245 \\     
 \hline
\end{tabular}
\end{table}
Figure \ref{fi:clustcoeff} presents time series plots of the mean total degree (which measures the total number of links that trader has), clustering coefficients (which measure the tendency of traders to establish transitive relationships) and assortativity coefficients (which measures the tendency of traders to interact with other traders that are similar to themselves) for the 201 networks in the NYMEX dataset.  These plots suggest the presence of at least a couple of change points in the structure of the network, including one around September 5, 2006 (which corresponds the date of introduction of electronic trading in this market via the CME Globex platform).  To investigate the presence of change points in more detail we fitted a Bayesian hidden Markov model with bivariate Gaussian emissions  (see Appendix \ref{ap:simpleHMM} for details on the model).  First we fitted the model to the bivariate time series of mean total degree and clustering coefficient, and then to the bivariate time series of clustering and assortativity coefficients.  Figure \ref{se:incidence_simpleHMM} shows the marginal posterior probability that any pair of weeks are assigned to the same latent state on each model.  These graphs illustrate that the analysis of networks based on summary statistics depends substantially on the ad-hoc choice of the summaries.  Indeed, although both graphs provide evidence of a change point around early September 2006, they disagree on whether other change points are present, and if so, when those happened.


\begin{figure}
\begin{center}
\includegraphics[scale=0.42,keepaspectratio]{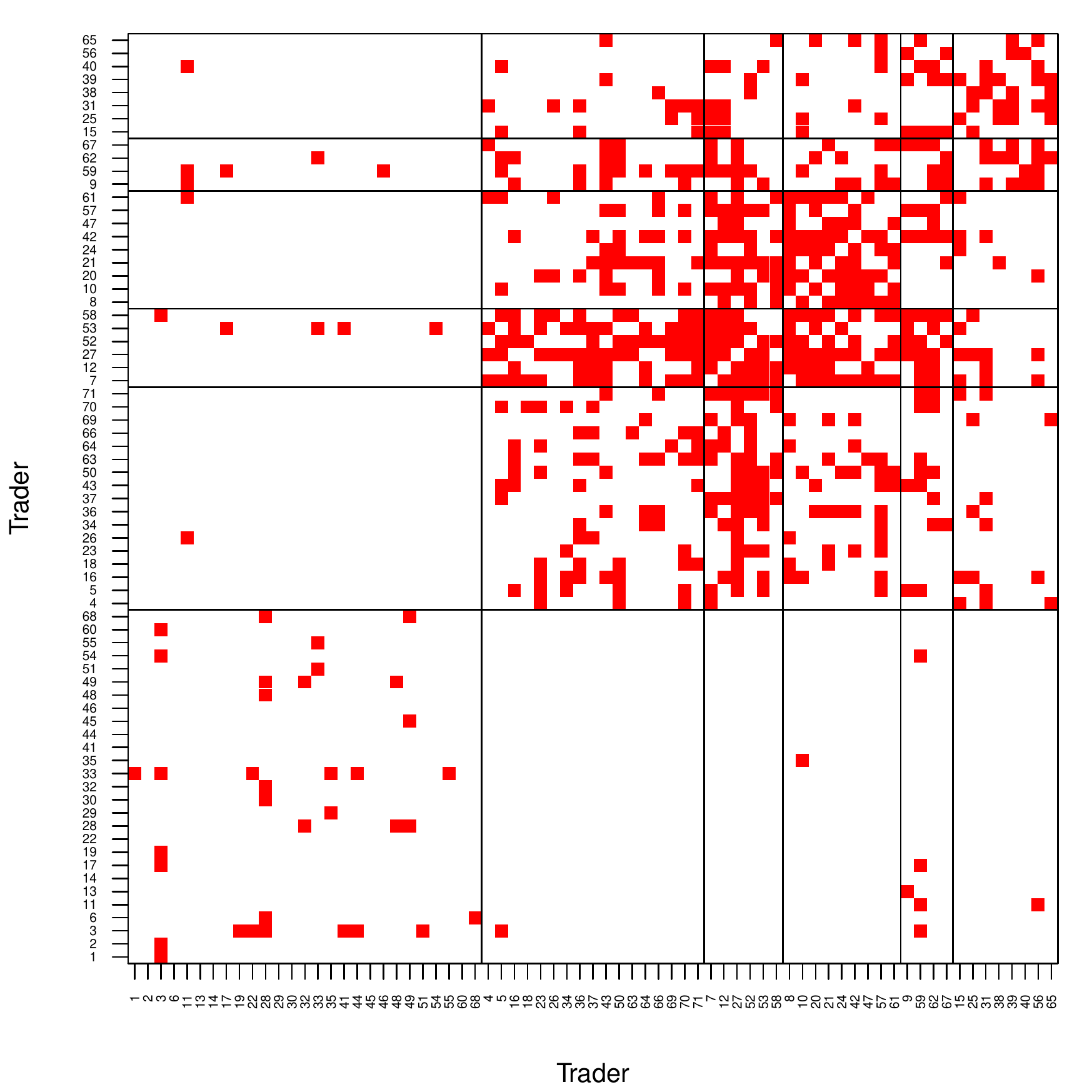}
\end{center}
\caption{Sociomatrix for the trading network associated with the week of February 22, 2005.  The solid lines suggest one possible partition of the traders into groups of structurally equivalent nodes.}\label{fi:single_network}
\end{figure}

Finally, figure \ref{fi:single_network} presents a matrix representation of the trading network associated with the week of February 22, 2005 (traders have been reordered to make the graph easier to read). The graph suggests the existence of groups of traders that are structurally equivalent, including a large group of inactive traders that do not participate on transactions during this particular week, as well as a couple of small group of traders with a high number of intra-group and a relatively low number of inter-group transactions.  This suggests that a stochastic blockmodel might be a reasonable model for individual trading networks.


\section{Modeling Approach}\label{se:modelingapproach}

\subsection{Stochastic blockmodels for financial trading networks}\label{se:singlenet}

We encode a financial trading network among $n$ traders using an $n \times n$ binary sociomatrix $\bfY = [ y_{i,j} ]$, where $y_{i,j} = 1$ if trader $i$ sold at least one contract to trader $j$, and $y_{i,j} = 0$ otherwise.  Since we focus on proprietary trading (i.e., transactions carried out by the traders with their own money, rather that their clients'), we adopt the convention $y_{i,i} \equiv 0$, as traders do not buy from themselves.  Note that treating the network as binary ignores information about the transactions such as the number, maturities, and prices of the contracts.  We proceed in this way for two reasons.  First, in some markets (i.e., black pools) the prices and number of contracts might not be disclosed, making it impossible to apply more general models.  Second, even if available, this extra information provides limit additional information about the identity of counterparties subject to contagion risks.  Nonetheless, the framework we describe here can be easily extended to more general types of weighted networks.

A stochastic blockmodel for $\bfY$ assumes that its entries are conditionally independent given two sets of parameters:  a vector of discrete indicators $\bfxi = (\xi_1, \ldots, \xi_n)$, where $\xi_i = k$ if and only if trader $i$ belongs to community $k = 1,\ldots K$, and a $K \times K$ matrix $\bfTheta = [\theta_{k,l}]$ such that $\theta_{k,l}$ represents the probability that a member of community $k$ sells a contract to a member of community $l$.  Therefore,
\begin{align*}
y_{i,j} \mid \bfxi, \bfTheta &\sim \Ber(\theta_{\xi_i, \xi_j})  .
\end{align*}
Note that $K$ represents the maximum potential number of trading communities allowed a priori.  A posteriori, the effective number of trading communities $K^{*}$ present in the sample could potentially be smaller than $K$.

A Bayesian formulation for this model is completed by eliciting prior distributions for $K$, $\bfxi$, and $\bfTheta$.  In the sequel we set $K=\infty$ and let the indicators be independent a priori where
\begin{align*}
\Pr(\xi_i = k \mid \bfw) &= w_k ,     &    k &= 1,2, \ldots ,
\end{align*}
and the vector of weights $\bfw = (w_1, w_2, \ldots)$ is constructed so that
\begin{align}\label{eq:PYP}
w_k &= v_k \prod_{s < k} \{ 1 - v_s \}  ,         &  v_k &\sim \bet(1 - \alpha, \beta + \alpha k),
\end{align}
for $0 \le \alpha < 1$ and $\beta > - \alpha$.  Note that, by setting $K = \infty$, the model allows for the effective number of components $K^{*}$ to be as large as the number of traders $n$, for any $n$.

The formulation in \eqref{eq:PYP} is equivalent to the constructive definition of the Poisson-Dirichlet process \citep{Pitman,Yor}, with $\alpha=0$ leading to the Dirichlet process.  Hence, the implied prior on the effective number of trading communities $K^{*}$ and the size of those communities, $m_1, \ldots, m_{K^{*}}$, is given by
$$
\frac{\Gamma(\beta+1)}{(\beta+ \alpha K^{*}) \Gamma(\beta + n)}  \prod_{k=1}^{K^{*}} (\beta + \alpha k) \frac{\Gamma(m_k - \alpha)}{\Gamma(1 - \alpha)}  .
$$
Note that larger values of $\alpha$ or $\beta$ favor a priori a larger effective number of $K^{*}$.  Setting $\alpha = 0$ leads to the prior expected number of communities to grow logarithmically with $n$, while for $\alpha > 0$ the expected number components grows as a power of the number of traders.

Consider now specifying a prior on the matrix of interaction probabilities $\bfTheta$.  In this case we let 
\begin{align*}
\theta_{k,l} &\mid a_O, b_O,  a_D, b_D \sim
\begin{cases}
   \bet(a_O, b_O)        &  k \ne l \\
   \bet(a_D, b_D)        &  k = l .
\end{cases}
\end{align*}

This prior is more general than those typically used in stochastic blockmodels, as it allows the distribution of the diagonal and off-diagonal elements of $\bfTheta$ to have different hyperparameters.  This ensures additional flexibility in terms of the implied degree distribution of the network, while still ensures that both $p(\bfY)$ and $p(\bfTheta)$ are jointly exchangeable, i.e., that the distributions are invariant to the order in which traders or communities are labeled \citep{Al81}.  In addition, it allows us to define an assortative index for the network as
\begin{align*}
\Upsilon & = \log \left\{ \E(\theta_{k,k} \mid a_D, b_D) \right\} - \log \left\{ \E(\theta_{k,l} \mid a_O, b_O) \right\} \\
& = \log \left\{  \frac{a_D}{a_D + b_D}  \right\}  -  \log \left\{  \frac{a_O}{a_O + b_O}  \right\} ,
\end{align*}
and a cycle-type transitivity index 
$$
\chi = \Pr(y_{i,j} = 1 \mid y_{j,k}=1, y_{k,i} = 1, a_O, b_O, a_D, b_D, \alpha, \beta) = \frac{\chi_N}{\chi_D}  ,
$$
where
\begin{multline*}
\chi_N =  \frac{(1-\alpha)(2-\alpha)}{(\beta+1)(\beta+2)} \frac{(a_D + 2)(a_D + 1)a_D}{(a_D + b_D + 2)(a_D + b_D + 1) (a_D+b_D)} \\
+  3\frac{(1-\alpha)(\beta+\alpha)}{(\beta+1)(\beta+2)} \frac{a_D}{(a_D + b_D)}  \left( \frac{a_O}{a_O+b_O} \right)^2  \\
+ \frac{(\beta+\alpha)(\beta+2\alpha)}{(\beta+1)(\beta+2)} \left(\frac{a_O}{a_O+b_O} \right)^3  ,
\end{multline*}
and
\begin{multline*}
\chi_D =  \frac{(1-\alpha)(2-\alpha)}{(\beta+1)(\beta+2)} \frac{(a_D + 1)a_D}{(a_D + b_D + 1) (a_D+b_D)} \\
+  2\frac{(1-\alpha)(\beta+\alpha)}{(\beta+1)(\beta+2)} \frac{a_D}{(a_D+b_D)}\frac{a_O}{(a_O+b_O)} \\
+ \frac{(\beta+\alpha)(\beta+\alpha+1)}{(\beta+1)(\beta+2)} \left( \frac{a_O}{ a_O+b_O} \right)^2 .
\end{multline*}
These two indexes are model-based alternatives to assortativity by degree and the clustering coefficients discussed in Figure \ref{fi:clustcoeff} \citep{RoRe13}.

\subsection{Hidden Markov models for time series of financial trading networks}\label{se:hmm_networks}

We are interested in extending the hierarchical blockmodel described in Section \ref{se:singlenet} to model a time series of financial trading networks $\bfY_1, \ldots \bfY_T$.  The extension is built with two goals in mind.  First, we are interested in identifying events associated with structural changes in the network and, therefore, in the microstructure of the market.  Second, we aim at making short-term predictions about the structure of the network in future periods.  For these reasons, we focus our attention on the use of hidden Markov models for network data.  Hidden Markov models are widely used in financial (e.g., see \citealp{RyTeAs98} and references therein) and biological (e.g., \citealp{YaPaRoHo11} and references therein) applications where there is interest in identifying structural changes in the system under study.  Hence, they represent a natural alternative in this context.

More specifically, consider now a sequence $\bfY_{1},\ldots,\bfY_{T}$ of binary trading networks observed over $T$ consecutive time intervals, where all networks are associated with a common set of $n$ traders.  In addition, let $\zeta_1, \ldots, \zeta_T$ be a sequence of unobserved state variables such that $\zeta_t = s$ indicates that the market is in state $s \in \{ 1, 2, \ldots, S \}$ during period $t = \{1, 2, \ldots, T \}$.  Each state has associated with it a vector of community indicators $\bfxi_s = (\xi_{1,s}, \ldots, \xi_{n,s})$ with $\xi_{i,s} \in \{1, 2, \ldots, K\}$ and a matrix of interaction probabilities $\bfTheta_{s} = [\theta_{k,l,s}]$ representing, respectively, the grouping of traders into trading communities and the probabilities of trades occurring between communities when the system is in state $s$.  Analogously to our previous discussion, $S$ and $K$ represent the maximum number of states and the maximum number of trading communities allowed by the model a priori.  A posteriori, the effective number of states $S^{*}$ and the effective number of communities on each state $K^{*}_{1}, \ldots, K^{*}_{S}$ is potentially smaller than $S$ and $K$, respectively.

Conditionally on the state parameters, observations are assumed to be independent, i.e.,
\begin{align*}
y_{i,j,t} \mid \zeta_t, \{ \bfxi_s \}, \{ \bfTheta_s \} \sim \Ber( y_{i,j,t} \mid \theta_{\xi_{i,\zeta_t}, \xi_{j, \zeta_t}, \zeta_t}).
\end{align*}
Hence, the joint likelihood for the data can be written as
\begin{align*}
p\left( \{ \bfY_t \} \mid \{ \bfzeta_t \}, \{ \bfxi_s \}, \{ \bfTheta_s \} \right) &= \prod_{t}^{T} \prod_{i=1}^{n} \prod_{\stackrel{j=1}{j \ne i}}^{n}
\theta_{\xi_{i,\zeta_t},\xi_{j, \zeta_t}, \zeta_t}^{y_{i,j,t}}  \left( 1 - \theta_{\xi_{i,\zeta_t},\xi_{j, \zeta_t}, \zeta_t} \right)^{1 - y_{i,j,t}}
\\
  &= \prod_{s=1}^{S} \prod_{k=1}^{K} \prod_{l=1}^{K} \prod_{(i,j,t) \in A_{k,l,s}} \theta_{k,l,s}^{y_{i,j,t}} \left( 1 - \theta_{k,l,s} \right)^{1 - y_{i,j,t}},
\end{align*}
where $A_{k,l,s}=\{(i,j,t): i \neq j, \zeta_t = s, \xi_{i,\zeta_t} =k, \xi_{j,\zeta_t}=l\}$ is the set of observations associated with the interactions between communities $k$ and $l$ in state $s$.

To account for the persistence in network structure illustrated in Figure \ref{fi:clustcoeff}, we assume that the evolution of the system indicators follows a first-order Markov process with transition probabilities
\begin{equation*}
p(\zeta_{t}=s \mid \zeta_{t-1}=r, \{ \bfpi_{r} \})= \pi_{r,s}  , 
\end{equation*}
where $\bfpi_r = (\pi_{r,1}, \ldots, \pi_{r,S})$, the $r$-th row of the transition matrix $\bfPi = [\pi_{r,s}]$, must satisfy $\sum_{s=1}^{S} \pi_{r,s} = 1$.  A natural prior for $\bfpi_r$ is a symmetric Dirichlet distribution,
\begin{align*}
\bfpi_r \mid \gamma & \sim \Dir\left( \frac{\gamma}{S}, \frac{\gamma}{S}, \ldots, \frac{\gamma}{S} \right) .
\end{align*}
Note that, as $S \to \infty$, the induced distribution of transitions over states is equivalent to that generated by a Dirichlet process prior with concentration parameter $\gamma$ (for example, see \citealp{GrRi01}).  Therefore the model is similar in spirit to the infinite hidden Markov model discussed in \cite{Ro10b} (see also \citealp{TeJoBeBl04}). However our construction does not couple the values of $\bfpi_1, \bfpi_2, \ldots$ through a common centering probability.  This is in contrast to the infinite hidden Markov model, where all transition probabilities into the same state are assumed to be similar.  Indeed, the structure of the infinite hidden Markov model implies that if state $s$ is highly persistent (i.e., $\pi_{s,s}$ is close to one), then the probability of transitioning from any other states into state $s$ will also tend to be large, a property that is unappealing when modeling financial trading networks.  Since $\gamma$ plays an important role in controlling the number of effective states $S^{*}$, its value is estimated from the data by assigning an exponential prior to it and carrying out a sensitivity analysis to evaluate the impact of our prior choice on model performance.

The specification of the model is completed by eliciting hierarchical priors on the state-specific parameters $\bfxi_1, \ldots, \bfxi_S$ and $\bfTheta_1, \ldots, \bfTheta_S$.  Following the discussion in Section \ref{se:singlenet}, we let
\begin{align*}
\Pr(\xi_{i,s}= k \mid \bfw_s) &=  w_{k,s} ,   &  k &= 1, 2,\cdot,
\end{align*}
where $w_{k,s}=v_{k,s} \prod_{h<k}\{ 1-v_{h,s} \}$ are weights constructed from a sequence $v_{1,s}, v_{2,s}, \ldots$ where $v_{k,s} \sim \bet(1-\alpha_s, \beta_s + k \alpha_s)$.  Again, since the hyperparameters $\alpha_s$ and $\beta_s$ play a critical role in controlling the number of expected trading communities, they are assigned independent hyperpriors $\alpha_s \sim p(\alpha_s)$
and $\beta_s \sim p(\beta_s)$.  A natural choice is to assign $\alpha_s$ a uniform prior on the unit interval and $\beta_s$ an exponential prior, while carrying out a sensitivity analysis that involves priors that favor small values of $\alpha_s$ as well as priors that favor both lower and higher values for $\beta_s$.


Similarly, the interaction probabilities are assigned priors
\begin{align*}
\theta_{k,l,s} &\mid a_{s,O}, b_{s,O},  a_{s,D}, b_{s,D} \sim
\begin{cases}
   \bet(a_{s,O}, b_{s,O})        &  k \ne l \\
   \bet(a_{s,D}, b_{s,D})        &  k = l .
\end{cases}
\end{align*}
where $\{ a_{s,O} \}$, $\{ b_{s,O} \}$, $\{ a_{s,D} \}$, and $\{ b_{s,D} \}$ are independent and gamma distributed with shape parameter $c$ and unknown rates $d_O$, $e_O$, $d_D$ and $e_D$, which are in turn assigned exponential priors with means $\lambda_d$ and $\lambda_e$.



\section{Computation}\label{se:computation}

%
%
%

The posterior distribution associated with our hidden Markov model for stochastic blockmodels is not analytically tractable. Therefore, we implemented a Markov chain Monte Carlo algorithm \citep{Casella} that simulates a dependent sequence of random draws from the target distribution.  Given initial values for the parameters, these are successively updated from their full conditional distributions. Standard Markov chain theory ensures that, after an appropriate burn-in, the values of the parameters generated by the algorithm are approximately distributed according to the posterior distribution.  To derive the algorithm, we rely on the fact that the joint posterior distribution can be factorized as
\begin{multline}\label{eq:posterior_fact}
p\left(\{ \bfTheta_s\} \mid \{ \bfxi_s \},\{ \zeta_t \}, \{ a_{s,O} \}, \{ b_{s,O} \}, \{ a_{s,D} \}, \{ b_{s,D} \}, \{ \bfY_t \} \right) \times \\
p\left(\{ \bfxi_s \},\{ \zeta_t \}, \{ a_{s,O} \}, \{ b_{s,O} \}, \{ a_{s,D} \}, \{ b_{s,D} \}, \right. \\ \left. d_O, e_O, d_D, e_D,  \{ \alpha_s \}, \{ \beta_s \}, \gamma \mid \{ \bfY_t \} \right)
\end{multline}
%

Since the values of $\theta_{k,l,s}$ are conditionally independent a posteriori given the observations, the indicators $\{ \zeta_t \}$ and $\{ \xi_{i,s} \}$, and the prior parameters $\{ a_{s,O} \}$, $\{ b_{s,O} \}$, $\{ a_{s,D} \}$ and $\{ b_{s,D} \}$, the first term in \eqref{eq:posterior_fact} is easy to sample from.  Furthermore, conditionally on the other parameters in the model, the state indicators $\zeta_1, \ldots, \zeta_T$ are sampled jointly using a forward-backward algorithm \citep{Rabiner86}, while the full conditional distribution for each collection of indicators $\xi_{1,s}, \ldots, \xi_{n,s}$ is sampled using a collapsed (marginal) Gibbs sampler \citep{Ne00}.  Details of the algorithm are discussed in Appendix \ref{se:appendixmcmc}.
%
%
%
%


Given a sample from the previous Markov chain Monte Carlo algorithm,
\begin{multline*}
\left( \{ \bfTheta_s^{(b)}\}, \{ \bfxi_s^{(b)} \},\{ \zeta_t^{(b)} \}, \{ a_{s,O}^{(b)} \}, \{ b_{s,O}^{(b)} \}, \{ a_{s,D}^{(b)} \}, \right.  \\ \left. \{ b_{s,D}^{(b)} \},  d_O^{(b)}, e_O^{(b)}, d_D^{(b)}, e_D^{(b)},  \{ \alpha_s^{(b)} \}, \{ \beta_s^{(b)} \}, \gamma^{(b)} \right),   \quad \quad    b=1, \ldots, B,
\end{multline*}
obtained after an appropriate burn-in period, point and interval estimates for model parameters can be easily obtained by computing the empirical mean and/or the empirical quantiles of the posterior distribution.  For example, posterior co-clustering probabilities, $\omega_{t,t'} = \Pr(\zeta_t = \zeta_{t'} \mid \{ \bfY_t \})$ can be estimated as
$$
\hat{\omega}_{t,t'} = \Pr(\zeta_t = \zeta_{t'} \mid \{ \bfY_t \}) \approx \frac{1}{B} \sum_{b=1}^{B} \mathbb{I}(\zeta_t^{(b)} = \zeta_{t'}^{(b)})  ,
$$
where $\mathbb{I}(\cdot)$ denotes the indicator function.  The estimates can be arranged into a co-clustering matrix $[\hat{\omega}_{t,t'}]$, which can in turn be used to identify the state of the system at each time period through a decision-theoretic approach (e.g., see \citealp{LaGr06}).  A similar procedure can be used to identify communities on each period.

The samples from the posterior distribution can also be used as the basis for prediction.  
For this purpose, note that the probability that trader $i$ sells at least one security to trader $j$ in the unobserved period $T+1$ can be estimated by
\begin{align*}
\E\left( y_{i,j,T+1} \mid \{ \bfY_t \} \right) \approx  \frac{1}{B} \sum_{b=1}^{B} \pi^{(b)}_{\zeta^{(b)}_T, s} \, \theta_{\xi^{(b)}_{i,s}, \xi^{(b)}_{j, s}} .
\end{align*}
Using a simple 0/1 utility function, a future sell trade from trader $i$ to trader $j$ is predicted as $\hat{y}_{i,j,T+1} = \mathbb{I}(\E\{ y_{i,j,T+1} \mid \{ \bfY_t \} \} >  f)$, for some threshold $f$ that reflects the relative cost associated false positive and false negative links.

\section{Analysis to the NYMEX natural gas futures market}\label{se:NYMEXillust}

In this section we analyze the sequence of $T=201$ weekly financial trading networks from transactions between 71 traders in the NYMEX natural gas futures market introduced in Section \ref{se:data}.  The results presented in this section are based on 100,000 iterations collected after a burn-in period of 10,000 iterations.  Convergence of the algorithm was diagnosed using the single-chain approach discussed in \cite{Ge92} and by a visual evaluation of trace plots.  We monitored the log-likelihood function, as well as the number of active states $S^{*}$ and the mean and variance over time of the assortativity and transitivity indexes $\{ \Upsilon_t \}$ and $\{ \chi_t \}$.  In terms of hyperparameters, the maximum number of states is set to $S=30$, the prior means for $\gamma$ and $\{ \beta_s \}$ are assigned exponential priors with unit mean, and the priors for $d_O$, $e_O$, $d_D$ and $e_D$ are exponential distributions with mean 2.  This specification implies that, a priori, $\E(\Upsilon_t) = 0$ for all $t=1, \ldots, T$, so that we favor neither assortative nor dissasortive trading communities a priori.
\begin{figure}
\begin{center}
\includegraphics[scale=0.35,keepaspectratio]{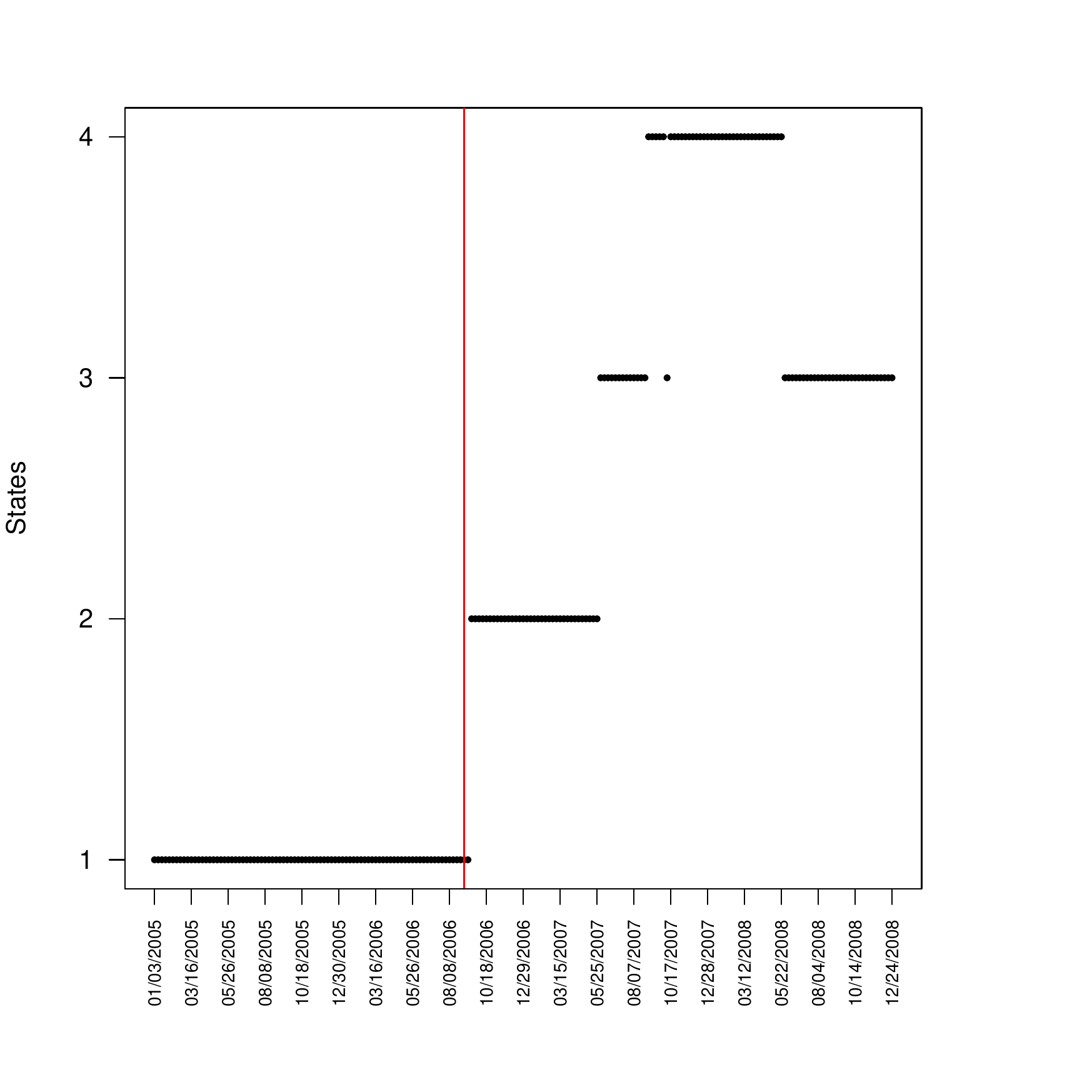}
\hspace{-0.5cm}
\includegraphics[scale=0.35,keepaspectratio]{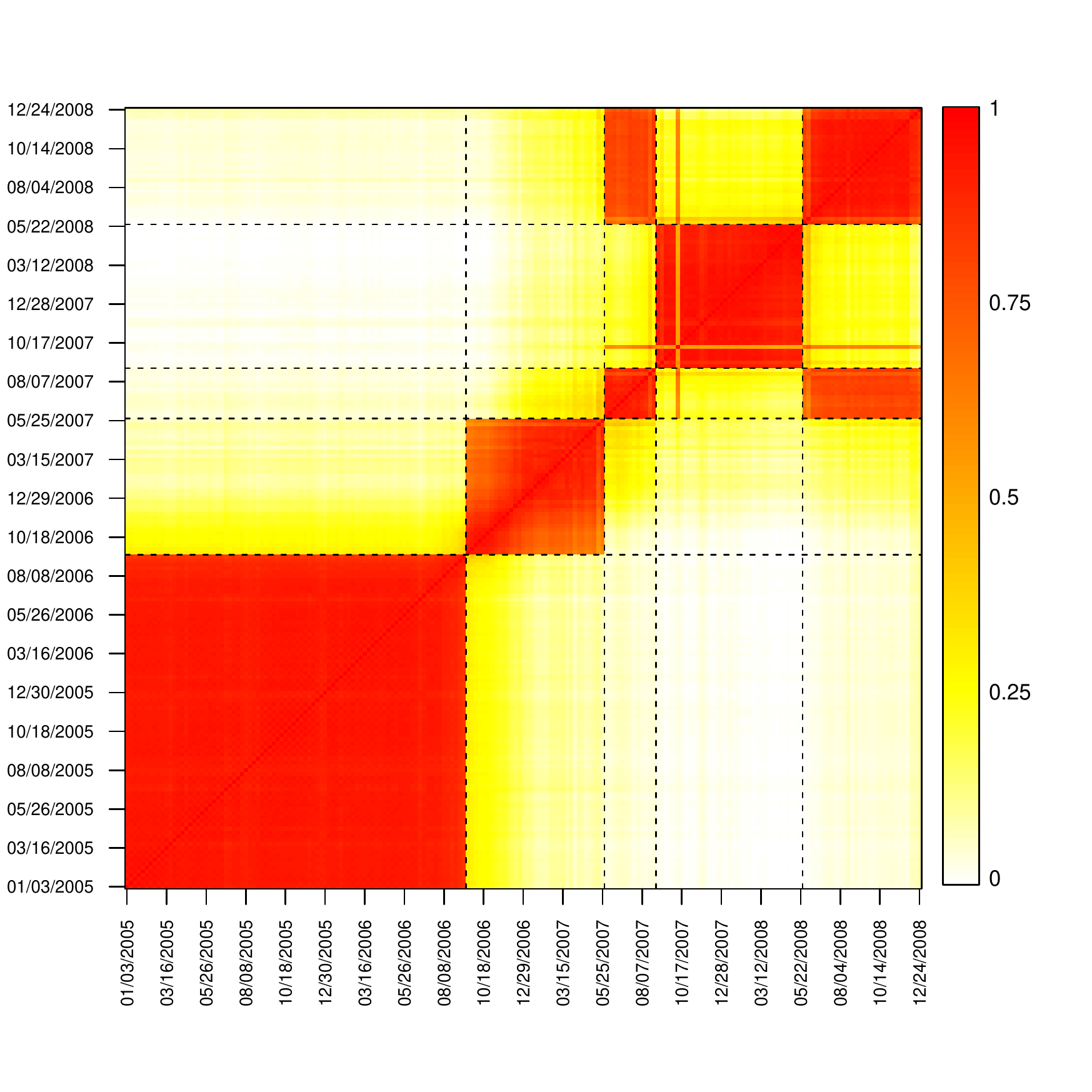}
\end{center}
\vspace{-0.5cm}
\caption{In the left panel, point estimate of the states for the 201 weeks observed for the trading network with the vertical line indicating the introduction of the electronic platform on week 85. On the right, mean posterior pairwise incidence matrix for the NYMEX networks under our blockmodel HMM, illustrating the uncertainty associated with this point estimate.}\label{fi:stateclustering}
\end{figure}

\vspace{-0.4cm}

\subsection{Identifying changes in market microstructure}

Figure \ref{fi:stateclustering} presents the posterior estimate of the co-clustering matrix for the latent states $\zeta_1, \ldots, \zeta_T$, along with a point estimator for the grouping of networks into states (recall Section \ref{se:computation}). This point estimator suggests that the structure of the trading networks alternates between four highly persistent states. The first state runs between early January 2005 and early September 2006, when the electronic market is introduced.  The second state runs between early September 2006 and early May 2007, when the system transitions to a new state for a short period of 3 months.  After that, the system seems to transition to a fourth state in early August 2007 (interestingly, the beginning of the recent financial crises), where it stays for 37 weeks before returning to the third state in early June 2008 (which coincides with some of the largest drops in the S\&P500 energy sector index over the last 13 years). Also, it is clear from the heatmap that, although there is some uncertainty associated with this point estimate of the system states (mostly in time of the transitions between states three and four), this uncertainty is relatively low.  Note that these results have some similarities with those we reported in Figure \ref{se:incidence_simpleHMM}, but also some important differences.  In particular, all models agree on the presence of a change point associated with the introduction of electronic trading on September 5, 2006, but disagree on the timing and structure of other change points.

\begin{figure}
\begin{center}
\includegraphics[scale=0.34,keepaspectratio]{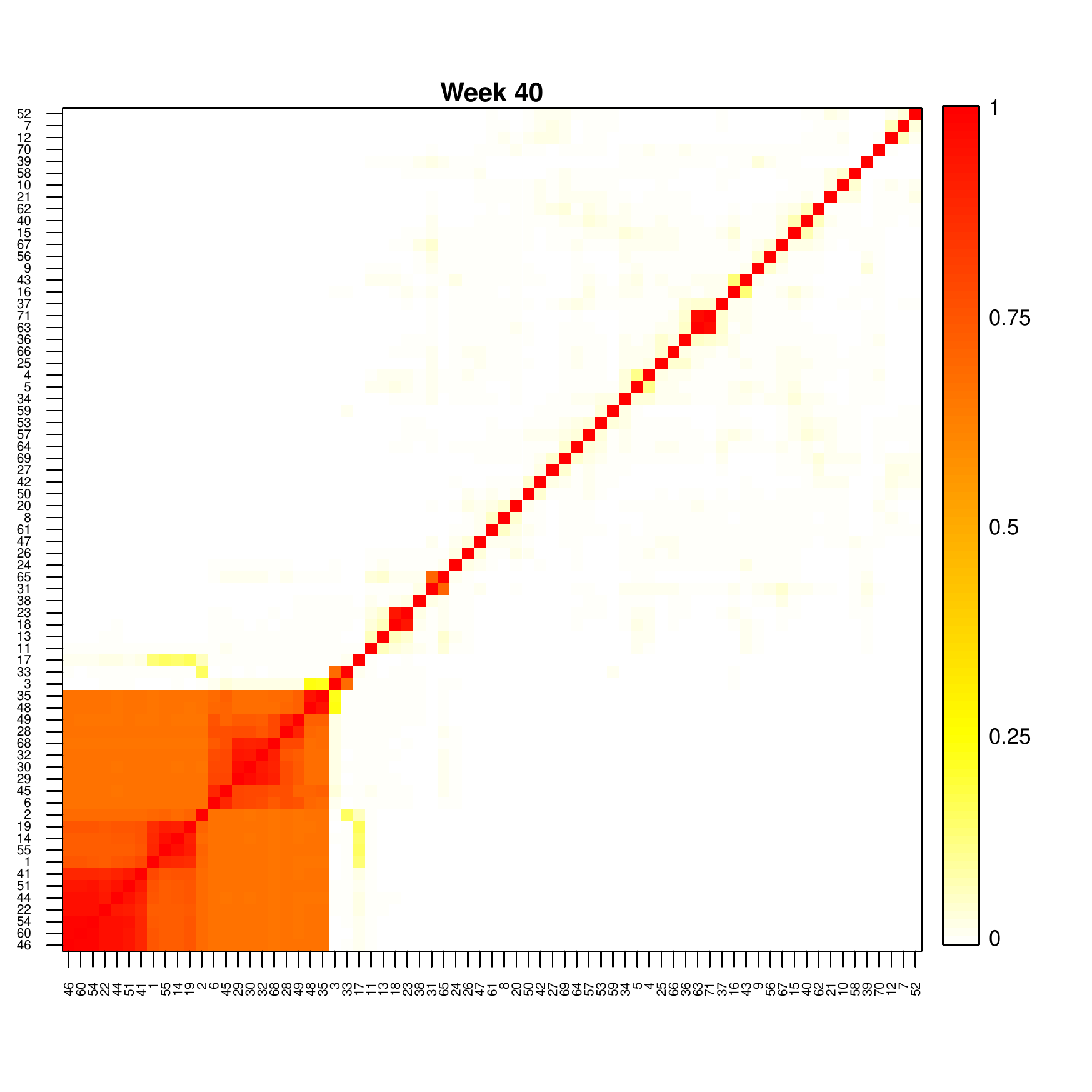}
\hspace{-0.4cm}
\includegraphics[scale=0.34,keepaspectratio]{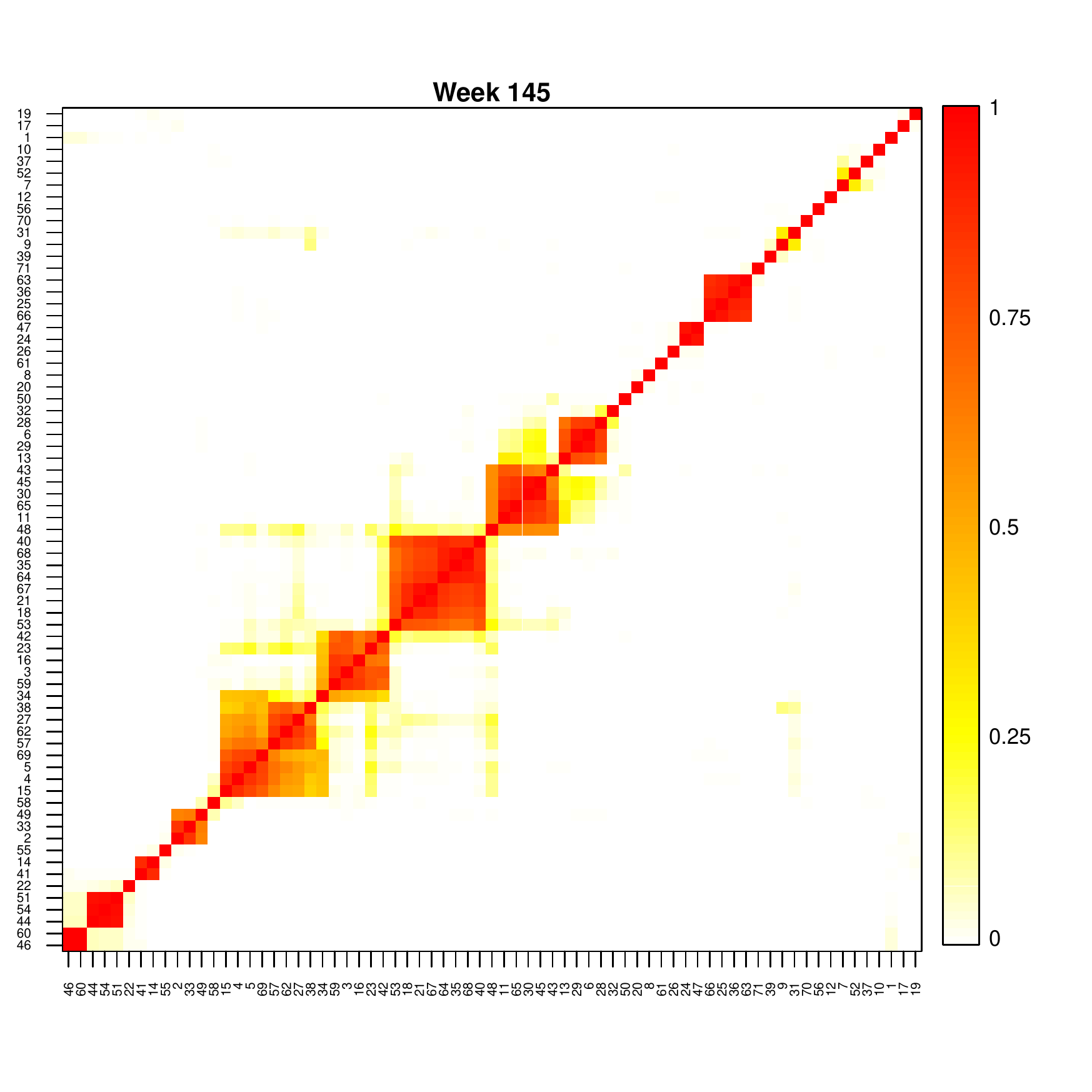}
\end{center}
\vspace{-0.5cm}
\caption{Mean posterior pairwise incidence matrices of traders for $t=40$ from state 1 and $t=145$ from state 4.}\label{fi:communitystructure}
\end{figure}

Figure \ref{fi:communitystructure} shows estimates of the community structure associated with two different weeks, that of October 11, 2005 ($t=40$) and that of November 14, 2007 ($t=145$).  We selected these dates because they are representative of states 1 and 4.  Note that, although there are some similarities, the overall structure of the communities is quite different.  State 1 is characterized by a large group of 25 mostly inactive traders, while all other traders tend to fall, for the most part, into singleton clusters.  On the other hand, while state 4 also exhibits a number of singleton clusters, it also shows a number of small communities comprising between 5 and 10 traders each.

\begin{figure}
\begin{center}
\includegraphics[scale=0.5,keepaspectratio]{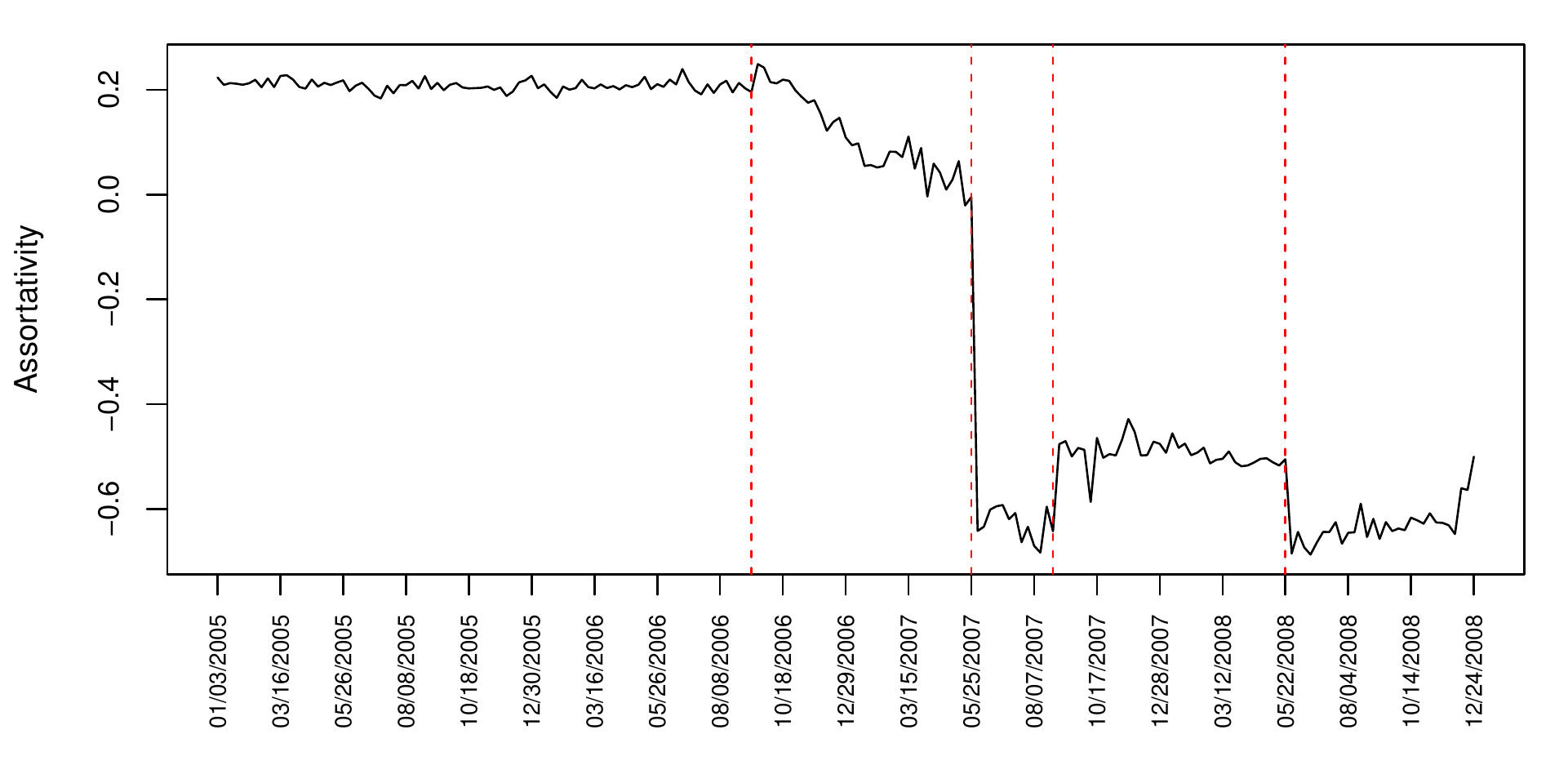}
\includegraphics[scale=0.5,keepaspectratio]{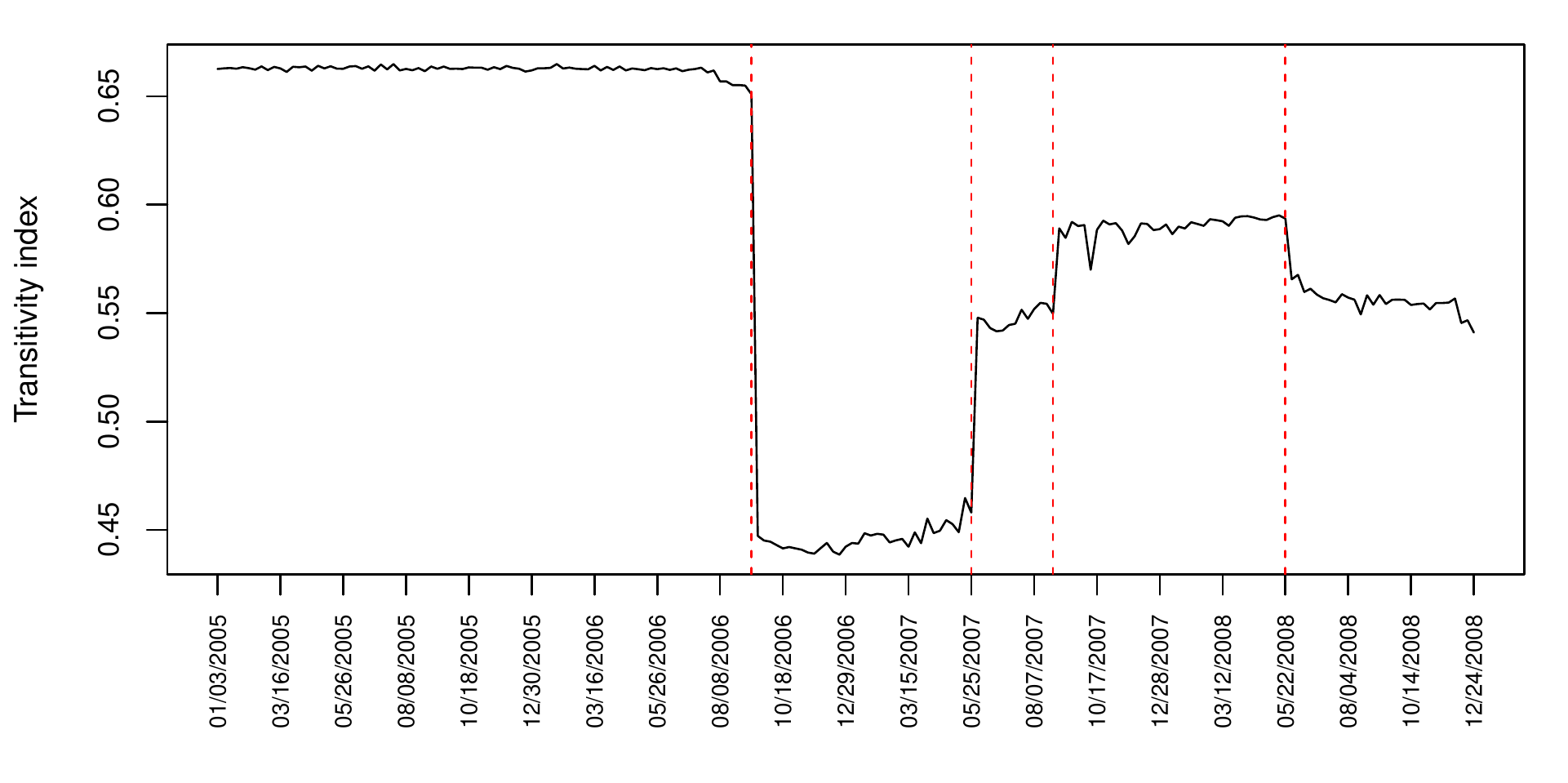}
\end{center}
\vspace{-0.5cm}
\caption{Time series plot for assortativity and transitivity indexes. The vertical line represents the transitions across states identified from Figure \ref{fi:stateclustering}.}\label{fig:assortindexmodel}
\end{figure}

Figure \ref{fig:assortindexmodel} shows time series plots for the estimates of the assortativity and transitivity indexes $\Upsilon_1, \ldots, \Upsilon_T$ and $\chi_1, \ldots, \chi_T$. Recall that these quantities are model-based alternatives to the assortativity by degree and the clustering coefficient presented in Figure \ref{fi:clustcoeff}.  Both sets of plots share some common features, revealing mild assortativity and higher transitivity before September 2006 and highly disassortative networks with lower transitivity afterwards.  This makes sense because we would expect that the introduction of an electronic market would limit the effect of social connections among traders (which tend to be assortative and transitive) and favor connections based on differential trending strategies (which tend to be disassortative).

Finally, Table \ref{ta:hyperparameters} shows point estimates and credible intervals associated with some hyperparameters in the model, both a priori and a posteriori.  In all cases, the posterior estimates appear to be more concentrated and be centered around different values than the prior.

\begin{table}
  \caption{\label{ta:hyperparameters}  Prior and posterior point estimates and credibility intervals for some model hyperparameters.}   \centering
\fbox{
\begin{tabular}{c|cc|cc}
\multirow{2}{*}{Parameter}  & Posterior & Posterior 95\%    &  Prior  &  Prior  95\%  \\
                                                 & mean       & credible interval  & mean &  credible interval  \\  \hline \hline
$\alpha_{40}$    & $0.748$ & $(0.198, 0.942)$  & $0.500$  &  $(0.025,0.975)$  \\
$\alpha_{100}$  & $0.524$ & $(0.240, 0.851)$  &  $0.500$  &  $(0.025,0.975)$  \\
$\alpha_{145}$  & $0.587$ & $(0.264, 0.785)$  &  $0.500$  &  $(0.025,0.975)$  \\    \hline
$\beta_{40}$    & $1.225$ & $(0.034, 4.526)$  & 1.000  &  (0.025, 3.689)  \\
$\beta_{100}$  & $2.518$ & $(0.107, 7.563)$  & 1.000  &  (0.025, 3.689)  \\
$\beta_{145}$  & $1.223$ & $(0.034, 4.510)$ & 1.000  &  (0.025, 3.689)  \\   \hline
$\gamma$  & $0.344$ & $(0.090, 0.814)$  &  1.000  &  (0.025, 3.689)  \\
\end{tabular}
}
\end{table}

%
%

\subsection{Network prediction}

As we discussed in the introduction, besides identifying change points in market microstructure, one of our goals is to predict future trading partnerships.  To assess the predictive capabilities of the model we ran an out-of-sample crossvalidation exercise where we held out the last ten weeks in the dataset and made one-step-ahead predictions for the structure of the held-out networks. More specifically, for each $t=191,192,\ldots,200$ we use the information contained in $\bfY_1, \ldots, \bfY_{t}$ to estimate the model parameters and obtain predictions for $\hat{\bfY}_{t+1}$ for different values of the threshold $f$.  Each of these predictions is compared against the observed network $\bfY_{t+1}$, the number of false and true positives is computed, and a receiver operating characteristic (ROC) curve is constructed. For comparison purposes, the same exercise was performed with a temporal Exponential Random Graph (tERGM). We used the \texttt{xergm} package in R to estimate the tERGM \citep{LeCranDes14}. 
More specifically, the tERGM is estimated with the \texttt{btergm} function, which implements the bootstrapped pseudolikelihood
procedure presented in \citet{DesCran12}. The model we fit includes all the typical ERGM terms, the square root of in and out-degrees as node covariates,
and the lagged network and the delayed reciprocity to model cross-temporal dependencies. The results show that the prediction ability of our model is good with an average AUC of 87\%. However, in this particular case the tERGM slightly but consistently outperforms our model, with an average AUC of 89\%.

\begin{figure}
\begin{center}
\includegraphics[scale=0.33,keepaspectratio]{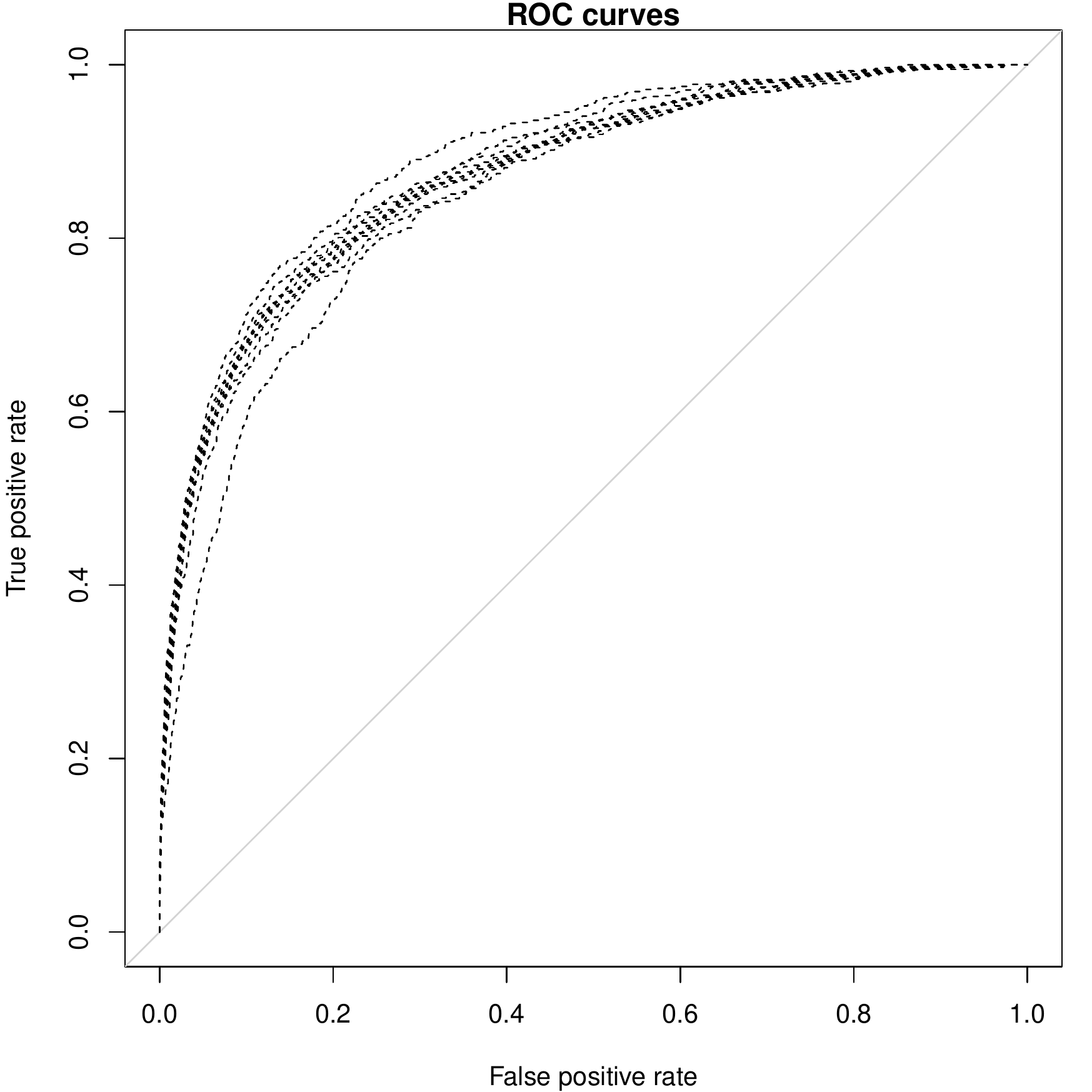}
\hspace{-0.3cm}
\includegraphics[scale=0.33,keepaspectratio]{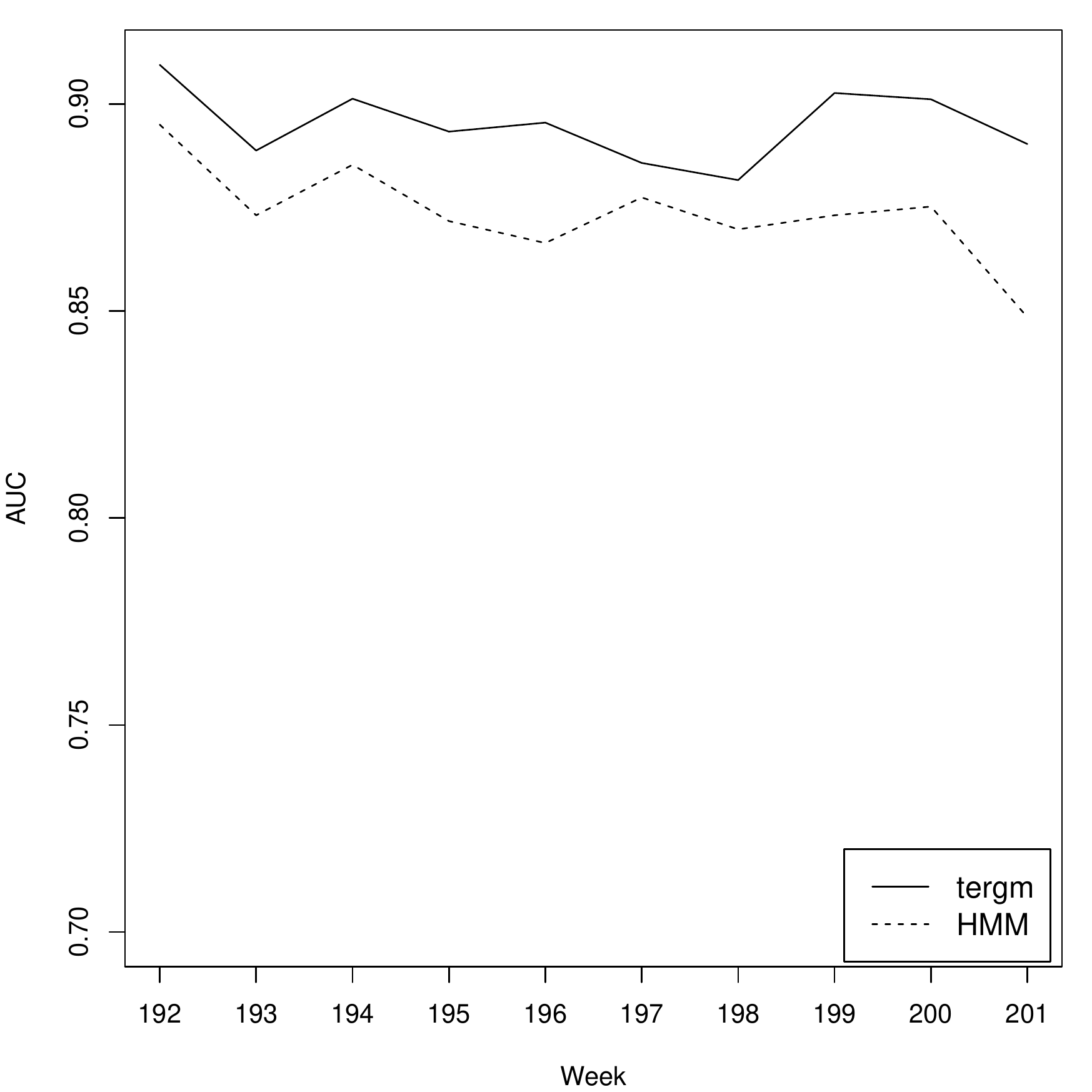}
\end{center}
\vspace{-0.5cm}
\caption{The left panel shows ten operating characteristic curves associated with one-step-ahead out of sample predictions from our hidden Markov model.  The right panel shows a time series plot of the Area Under the curves (AUC) for the tERGM and our proposed model (HMM).}\label{fig:rocandacu}
\end{figure}


\vspace{-0.3cm}

\subsection{Sensitivity analysis}

To assess the effect of our prior choice on posterior inference we conducted a sensitivity analysis where the model was fitted with somewhat different priors.  In particular, we used independent Beta priors with mean $1/10$ and variance $9/1100$ for each $\alpha_s$, as well as exponential priors with means $1/3$ and $3$ for each $\beta_s$.  On the other hand, a exponential priors with mean $2$ were also used for $d_O$, $e_O$, $d_D$ and $e_D$.  Although inferences on the community structure were somewhat affected by prior choices, inferences on the state parameters as well as the assortativity and transitivity indexes and the predictive performance were essentially unchanged.

\section{Discussion\label{se:discussion}}

We have presented a class of hidden Markov models for financial trading networks that have clear potential for market regulatory oversight.  Key application of these models include identifying specific events (such as large trader failures or specific changes in market rules) that affect market stability, as well as identifying frequent trading counterparties that might be likely collusion partners or particularly at risk in case of bankruptcies.

In this  paper we have focused on models for binary networks where only the presence/absence of transactions over a week is recorded.  However, when information about volumes is available, the model can be easily extended to incorporate this information.

Although the use of a hidden Markov model allows us to account for time dependence and is useful for identifying structural changes in the system, a structure that assumes abrupt changes in the network might be too restrictive for predictive purpose.  In the future we plan to evaluate models based on fragmentations and coagulations (e.g., see \citealp{Be06}) that allow for smooth evolution in the community structure, as well as extensions of auto logistic models that might allow for improved predictions.

\appendix

\small
\section{Hidden Markov model with bivariate normal emissions}\label{ap:simpleHMM}

In section \ref{se:data} we fit a hidden Markov model with bivariate Gaussian emissions for different pairs of summary statistics on the NYMEX network.  In this Appendix we provide a detailed formulation of the model.

Let $\bfx_t = (x_{1,t}, x_{2,t})'$, where $x_{1,t}$ and $x_{2,t}$ are two summary statistics (such as the clustering and assortativity coefficients) of the network observed on week $t$.  We assume that
\begin{align*}
\bfx_{t} \mid \zeta_t^{*}, \{ \bfmu_s \}, \{ \bfOmega_s \}  &\sim \normal \left( \bfmu_{\zeta^{*}_t}, \bfOmega_{\zeta^{*}_t} \right)    ,    &  t &= 1,\ldots, T ,
\end{align*}
where $\bfmu_s \sim \normal\left( \mathbf{d}, \mathbf{D} \right)$ and $\bfOmega_s \sim \IWis \left(a, \mathbf{B} \right)$ independently for each $s=1, \ldots, R$ and, as in our other model in this paper, the state indicators satisfy
\begin{align*}
p\left(\zeta_t^{*} = s \mid \zeta_{t-1}^{*} = r, \{ \bfpi^{*}_r \} \right) &= \pi^{*}_{r,s}    ,    &    \bfpi_r \mid \gamma^{*} &\sim \Dir \left( \frac{\gamma^*}{R}, \ldots, \frac{\gamma^*}{R} \right) .
\end{align*}
For the analysis shown in Section \ref{se:data} we set $\gamma^{*} = 1$, and set $\mathbf{d}$ to the mean and $\mathbf{D}$ and $\mathbf{B}$ both to the variance-covariance matrix of the observations. The estimates of the pairwise probabilities $\Pr(\zeta^{*}_t = \zeta^{*}_{t'} \mid \{ \bfx_t \})$ were obtained from 10,000 iterations (obtained after a burn-in period of 1,000 samples) of an MCMC algorithm that alternates through sampling $\{ \bfmu_s \}$, $\{ \bfOmega_s \}$ and $\{ \bfzeta^{*}_t \}$ from their corresponding full conditional posterior distributions.  The details of the algorithm are very similar to the one discussed in Appendix \ref{se:appendixmcmc} for the hidden Markov model with blockmodel emissions.

\section{Details of the computational algorithm for the hidden Markov model with blockmodel emissions}\label{se:appendixmcmc}

Here, we provide the details of the MCMC algorithm discussed in \ref{se:computation}.  The algorithm proceeds by updating the model parameters from the following full conditional distributions:

\begin{description}
  \item[(a)] For each $i=1,\ldots,n$ and occupied states $s$, $\xi_{i,s}=k$ with probability
 \begin{align*}
\Pr(\xi_{i,s}=k \mid \cdots, \bfY)
\end{align*}
\begin{align*}
=\left\{  \begin{array}{ll}
%
    (m_{k}^{-i}-\alpha_s)\prod\limits_{l=1}^{K^{*}_{s,-i}}\frac{p( \{ y_{i,j,t} : (i,j,t) \in A_{k,l,s}^{i} \})}{p( \{ y_{i,j,t} : (i,j,t) \in A_{k,l,s}^{-i}\} )}
    \frac{p( \{ y_{j,i,t} : (i,j,t) \in A_{k,l,s}^{i} \})}{p( \{ y_{j,i,t} : (i,j,t) \in A_{k,l,s}^{-i}\} )}, & \hbox{$k\leq K^{*}_{s,-i}$} \\\\
    (\beta_s+\alpha_{s}K^{*}_{s,-i})\prod\limits_{l=1}^{K^{*}_{s,-i}}p( \{ y_{i,j,t} : (j,t) \in A_{l,s}^{-i}\} )   & \\
    \;\;\;\;\;\;\;\;\;\;\;\;\;\;\;\;\;\;\;\;\;\;\;\;\;\;\;\;\;\;\;\;\;\;\;\;\;\;\;\;\;\;\;\;\;\;\;\;\;\;\;\; p(\{ y_{j,i,t} : (j,t) \in A_{l,s}^{-i}\} ), & \hbox{$k=K^{*}_{s,-i}+1$,}
  \end{array}
\right.
\end{align*}
where $K^{*}_{s,-i}=\max_{j\neq i}\{\xi_{j,s}\}$, $m_{k}^{-i}=\sum_{j\neq i} \mathbb{I}_{(\xi_{j,s}=k)}$, 
\begin{align*}
A_{k,l,s}^{-i} &= \{(i',j',t):i'\neq j'\neq i, \zeta_{t}=s, \xi_{i',\zeta_{t}}=k, \xi_{j',\zeta_{t}}=l\} , \\
A_{k,l,s}^{i} &= \{(i',j',t):i'=i, \zeta_{t}=s, \xi_{j',\zeta_{t}}=l\}\bigcup A_{k,l,s}^{-i},  \\
A_{l,s}^{-i} &= \{(j,t):  j\neq i, \zeta_{t}=s, \xi_{j,\zeta_{t}}=l\} ,
\end{align*}
and the marginal predictive distribution, $p( \{ y_{i,j,t} : (i,j,t) \in A \} )$ is given by
\begin{align*}
\dfrac{\Gamma(\sum_{A}y_{i,j,t}+a_s) \Gamma(|A|+b_s-\sum_{A}y_{i,j,t})}{\Gamma(a_s+b_s+|A|)}
\dfrac{\Gamma(a_s+b_s)}{\Gamma(a_s)\Gamma(b_s)}.
\end{align*}
and $|A|$ is the number of elements in $A$.

  \item[(b)] Since the prior for $\theta_{k,l,s}$ is conditionally conjugate, we update these
  parameters for $k,l \in \{1,\ldots,K^{*}_s \}$ by sampling from
  \begin{align*}
\theta_{k,l,s}\mid \cdots,\bfY \sim \text{Beta}
\left(\sum_{A_{k,l,s}}y_{i,j,t}+a_s,m_{k,l,s}+b_s-\sum_{A_{k,l,s}}y_{i,j,t}\right)
\end{align*}
for  $A_{k,l,s}=\{(i,j,t): i \neq j, \zeta_t = s, \xi_{i,\zeta_t} =k, \xi_{j,\zeta_t}=l\}$ and $m_{k,l,s}=|A_{k,l,s}|$.

  \item[(c)] Since the prior for the transition probabilities is conditionally
  conjugate, the posterior full conditional for $\bfpi_r$, $r=1,\ldots,S$ is the Dirichlet distribution
\begin{equation*}
p(\bfpi_r \mid \cdots, \bfY)=\prod\limits_{s=1}^{S}\displaystyle\pi_{r,s}^{\gamma/S+n_{rs}-1}
\end{equation*}
for $n_{rs}=|\{t: \zeta_{t-1}=r, \zeta_{t}=s\}|$.

\item[(d)] The posterior full conditional of $\gamma$ is
\begin{align*}
p(\gamma \mid \cdots,\bfY) \propto p(\gamma)\displaystyle\prod\limits_{s=1}^{S}\dfrac{\Gamma(\gamma)}{\Gamma(\gamma+n_s)}\gamma^{L_s}
\end{align*}
where $n_s=|\{t: \zeta_t=s\}|$ and $L_s=\sum_{r} \mathbb{I}_{n_{s,r}>0}$ for
$n_{s,r}=|\{t: \zeta_{t-1}=s,\zeta_{t}=r \}|$.
Since this distribution has no standard form, we update $\gamma$ using a random walk Metropolis-Hastings
algorithm with symmetric log-normal proposal,
\begin{align*}
\log\{\gamma^{(p)}\} \mid \gamma^{(c)} \sim \normal \left(\log\{\gamma^{(c)}\},\kappa^{2}_{\gamma}\right)
\end{align*}
where $\kappa^{2}_{\gamma}$ is a tuning parameter chosen to get an average acceptance rate between $30\%$ and $40\%$ .

  \item[(e)]The posterior full conditional of the pairs $(a_{s,O},b_{s,O})$ and $(a_{s,D},b_{s,D})$ has the following general form:
\begin{align*}
p(a_s,b_s \mid \cdots,\bfY) \propto p(a_s \mid d)p(b_s \mid e)\prod\limits_{k=1}^{S}\prod\limits_{l=1}^{S}p(y_{i,j,t}\mid A_{k,l,s},m_{k,l,s} )
\end{align*}
for the marginal predictive $p(y_{i,j,t}\mid A_{k,l,s},m_{k,l,s})$ as defined in step (b), $A_{k,l,s}=\{(i,j,t): i \neq j, \zeta_t = s, \xi_{i,\zeta_t} =k, \xi_{j,\zeta_t}=l\}$ and $m_{k,l,s}=|A_{k,l,s}|$.
Since no direct sampler is available for this distribution, we update each pair using a random walk Metropolis-Hastings
algorithm with bivariate log-normal proposals,
\begin{multline*}
\left(\log\{a_{s}^{(p)}\},\log\{b_{s}^{(p)}\}\right)^{t} \mid \left(a_{s}^{(c)},b_{s}^{(c)}\right)^{t}  \sim \\ \normal \left[\left(\log\{a_{s}^{(c)}\},\log\{b_{s}^{(c)}\}\right)^{t} ,\bfSigma_{ab}\right]
\end{multline*}
where $\bfSigma_{ab}$ is a tuning parameter matrix chosen independently for diagonal and off-diagonal pairs of parameters.

\item[(f)] The parameters of the Poisson-Dirichlet process $(\alpha_s,\beta_s)$
   can be jointly updated using the algorithm described in \citet{Escobar}.

\item[(g)] The posterior full conditional distributions for the hyperparameters $d_O, e_O, d_D$, and $e_D$
correspond to gamma distributions with shape parameter $(cS^{*}+1)$ and rate parameters $(\sum_{S^{*}}a_{s,O}+\lambda_d)$,
$(\sum_{S^{*}}b_{s,O}+\lambda_e)$, $(\sum_{S^{*}}a_{s,D}+\lambda_d)$, $(\sum_{S^{*}}b_{s,D}+\lambda_e)$, respectively.

\end{description}

\bibliographystyle{plainnat}    
\bibliography{networks}

\end{document}